\documentclass[twocolumn]{autart}  

\usepackage[longnamesfirst]{natbib}
\usepackage{amsmath}
\usepackage{amssymb}
\usepackage{accents}

\newtheorem{thm1}{\bf Theorem}
\newtheorem{cor1}{\bf Corollary}
\newtheorem{lem1}{\bf Lemma}
\newtheorem{defn1}{\bf Definition}
\newtheorem{assum1}{\bf Assumption}
\newtheorem{rem1}{\bf Remark}

\newtheorem{prop1}{\bf Proposition}

\newcommand{\R}{\mathbb{R}}
\newcommand{\xx}{\mathsf{x}}
\newcommand{\vv}{\mathsf{v}}
\newcommand{\yy}{\mathsf{y}}
\newcommand{\uu}{\mathsf{u}}

\newcommand{\LL}{\mathcal{L}}
\newcommand{\NN}{\mathcal{N}}
\newcommand{\VV}{\mathcal{V}}

\newcommand{\UU}{\mathcal{U}}
\newcommand{\ZZ}{\mathcal{Z}}

\newcommand{\AAA}{\mathcal{A}}
\newcommand{\DDD}{\mathcal{D}}

\newcommand*{\QEDB}{\hfill\ensuremath{\square}}%

\newcommand{\oo}[1]{\accentset{o}{#1}}

\begin{document}
	
\begin{frontmatter}

\title{A Tool for Analysis and Synthesis of Heterogeneous Multi-agent Systems under Rank-deficient Coupling\thanksref{footnoteinfo}}

\thanks[footnoteinfo]{This work was supported by the National Research Foundation of Korea (NRF) grant funded by the Korea government (Ministry of Science and ICT) (No.~NRF-2017R1E1A1A03070342).
This work was done while Jin Gyu Lee is with Seoul National University.
This is a preprint of the following paper: Jin Gyu Lee and Hyungbo Shim, ``A tool for analysis and synthesis of heterogeneous multi-agent systems under rank-deficient coupling,'' published in Automatica, 2020, Elsevier reproduced with permission of Elsevier.
The final authenticated version is available online at: http://dx.doi.org/10.1016/j.automatica.2020.108952}

\author[Cambridge]{Jin Gyu Lee}\ead{jgl46@cam.ac.uk} $\;$ and $\;$
\author[SNU]{Hyungbo Shim}\ead{hshim@snu.ac.kr}
\address[Cambridge]{Control Group, Department of Engineering, University of Cambridge, Cambridge, United Kingdom.}
\address[SNU]{ASRI, Department of Electrical and Computer Engineering, Seoul National University, Seoul, Korea.}

		
\begin{abstract}
The behavior of heterogeneous multi-agent systems is studied when the coupling matrices are possibly all different and/or singular, that is, its rank is less than the system dimension.
Rank-deficient coupling allows exchange of limited state information, which is suitable for the study of multi-agent systems under output coupling.
We present a coordinate change that transforms the heterogeneous multi-agent system into a singularly perturbed form.
The slow dynamics is still a reduced-order multi-agent system consisting of a weighted average of the vector fields of all agents, and some sub-dynamics of agents.
The weighted average is an emergent dynamics, which we call a blended dynamics.
By analyzing or synthesizing the blended dynamics, one can predict or design the behavior of a heterogeneous multi-agent system when the coupling gain is sufficiently large.
For this result, stability of the blended dynamics is required.
Since stability of the individual agent is not asked, the stability of the blended dynamics is the outcome of trading off the stability among the agents.
It can be seen that, under the stability of the blended dynamics, the initial conditions of the individual agents are forgotten as time goes on, and thus, the behavior of the synthesized multi-agent system is initialization-free and is suitable for plug-and-play operation.
As a showcase, we apply the proposed tool to four application problems; distributed state estimation for linear systems, practical synchronization of heterogeneous Van der Pol oscillators, estimation of the number of nodes in a network, and a problem of distributed optimization.
\end{abstract}
		
\end{frontmatter}

\section{Introduction}
This paper studies the behavior of a multi-agent system whose individual agent dynamics is given by
\begin{subequations}\label{eq:eachdyn}\vspace{-2mm}
\begin{equation}\label{eq:eachdyna}
	\dot x_i = f_i(t, x_i) + k B_i \sum_{j \in \NN_i} \alpha_{ij} \left(x_j-x_i\right), \quad i \in \NN,
\end{equation}
or
\begin{equation}\label{eq:eachdynb}\vspace{-2mm}
	\dot x_i = f_i(t, x_i) + k \sum_{j \in \NN_i} \alpha_{ij} \left(C_j x_j - C_i x_i\right), \quad i \in \NN,
\end{equation}
\end{subequations}
where $\NN := \{1, \cdots, N\}$ is the set of agent indices with the number of agents $N$, and $\NN_i$ is a subset of $\NN$ whose elements are the indices of the agents that send the information to agent $i$.
The coefficient $\alpha_{ij}$ is the $ij$-th element of the adjacency matrix that represents the interconnection graph.
We are particularly interested in the case when the coupling gain $k \in \R_{>0}$ is sufficiently large.
In the description, the internal state of an individual agent is represented by $x_i \in \R^n$.
It is a {\em heterogeneous} multi-agent system in the sense that the vector field $f_i: \R \times \R^n \to \R^n$ and the coupling matrix $B_i \in \R^{n \times n}$ or $C_i \in \R^{n \times n}$ are possibly different from each other.
Note that the time-varying $f_i$ can include external inputs, disturbances, and/or noises, which are possibly different for different agents; for example, $f_i(t,x_i) = g_i(x_i,u_i(t),d_i(t))$, where $g_i$ is some vector field, $u_i$ is a control input, and $d_i$ is an external disturbance to agent $i$.
The vector field $f_i$ is assumed to be piecewise continuous in $t$, continuously differentiable with respect to $x_i$, locally Lipschitz with respect to $x_i$ uniformly in $t$, and $f_i(t, 0)$ is uniformly bounded for $t$.

Of particular interest in the current study is the case when the matrix $B_i \in \R^{n \times n}$ or $C_i \in \R^{n \times n}$ is symmetric and positive semi-definite,\footnote{If the $B_i$'s are not all symmetric but have non-negative real eigenvalues and if there exists a real nonsingular matrix $\Theta$ such that $\Theta B_i \Theta^{-1}$ are symmetric for all $i \in \NN$, then, by re-defining the state $x_i' := \Theta x_i$ and the vector field, it can be seen that the $x_i'$-dynamics satisfies the assumptions under consideration. The same applies to the case of $C_i$. An example is given in Section \ref{sec:application2}.} 
which we call a {\em rank-deficient} coupling matrix.
It is a relaxation of the case of the identity coupling matrix, which frequently appears in the literature.
This relaxation becomes useful when only some of the elements in the internal state $x_i$ are communicated and/or only a part of the integration of $x_i$ is affected by other agents.
A few examples of this utility are found in Section \ref{sec:application}.

{\bf Note for reading:}
In this paper, two different classes of systems, \eqref{eq:eachdyna} and \eqref{eq:eachdynb}, are studied concurrently in order to save space.
To avoid confusion, we ask the reader to first choose the system \eqref{eq:eachdyna} or \eqref{eq:eachdynb} depending on his/her interest, and then apply the following convention.
When the equation number is, for example, \eqref{eq:eachdyn}, it should be interpreted as \eqref{eq:eachdyna} or \eqref{eq:eachdynb}, depending on the reader's choice.
When the reader encounters two equations having numbers such as \eqref{eq:Bi} and \eqref{eq:Ci}, he/she should just focus on the one equation whose number ends with the letter (``a'' or ``b'') that corresponds to the reader's choice, and ignore the other equation.

Our first goal is to analyze the behavior of the heterogeneous multi-agent system \eqref{eq:eachdyn} without solving \eqref{eq:eachdyn}.
The approach is to consider the extreme case when $k \to \infty$, because it gives a clue to approximate the behavior of \eqref{eq:eachdyn} when $k$ is finite but sufficiently large.
For this, let us define a notion of the ``limiting solution.''

\begin{defn1}\label{def:limitingsolution}
The system \eqref{eq:eachdyn} with initial conditions $x_i(0)$, $i \in \NN$, is said to admit the {\em limiting solution} $\xi_i$, $i \in \NN$, if there is a continuous function $\xi_i: (0,\infty) \to \R^n$ such that
	\begin{equation}\label{eq:pwconv}
		\lim_{k \to \infty} x_i(t,k) = \xi_i(t), \quad \forall i \in \NN, \quad \forall t > 0,
	\end{equation}
where $x_i(\cdot,k)$ represents the solution $x_i(\cdot)$ of agent $i$ in \eqref{eq:eachdyn} with the coupling gain $k$ (and $x_i(0,k)=x_i(0)$).
Moreover, if for each $\eta > 0$, there exist $T > 0$ and $k^*$ such that, for each $k > k^*$, we have
$$\|x_i(t, k) - \xi_i(t)\| \le \eta, \quad \forall i \in \mathcal{N}, \quad \forall t \ge T,$$
then we say that the convergence is {\em uniform in time}.
\end{defn1}

It should be noted that the convergence in \eqref{eq:pwconv} is point-wise in time $t$; i.e., the minimal $k$ to make the convergence error at time $t$ less than a given number may depend on $t$.
Since the gain $k$ is finite and fixed over time in practice and we are often interested in the behavior for an infinite time horizon, the notion of the limiting solution is not very useful unless the convergence is uniform in time.
In the sequel, by way of the singular perturbation analysis, we introduce the so-called ``blended dynamics,'' which is independent of $k$, and we claim that, if the blended dynamics is complete, i.e., its solution exists for all $t>0$, then the system \eqref{eq:eachdyn} admits the limiting solution.
Moreover, we will prove that, if the blended dynamics is stable in a certain sense, then the convergence becomes uniform in time.
It is also shown that the limiting solution $\xi_i$ is a linear transformation of the solution to the blended dynamics.
Knowledge of $\xi_i$ is useful because one can predict the behavior of the multi-agent system at least approximately when $k$ is large, and the approximation error becomes arbitrarily small as $k$ becomes larger.
Motivated by this, one can {\em design} the blended dynamics first such that the trajectory $\xi_i$ behaves as desired, and then, synthesize a multi-agent system such that it has the desired blended dynamics.
We will show how this can be done in Section \ref{sec:application}.

Blended dynamics is in general a reduced-order multi-agent system consisting of a weighted average of individual vector fields $f_i$ and each agent's sub-dynamics.
The averaged part can be considered as a blend of all $f_i$'s, from which its name is derived.
This notion has already appeared in \citep{kim2012practical} for scalar linear systems, and in \citep{kim2013robustness,kim2016robustness} for scalar nonlinear systems under the terminology ``averaged dynamics.''
More recently, \cite{panteley2015practicalACC} introduced a notion called ``emergent dynamics'' (see also \citep{panteley2017synchronization}), which is, however, different from the blended dynamics in the sense that it is not a multi-agent system.
Moreover, they require stability of zero-dynamics in each agent and they take the average of the zero-dynamics when the emergent dynamics is constructed.
As a result, an arbitrarily high-precision approximate is not obtained for the behavior of the multi-agent system.
In spite of different names and notions, there is a common philosophy in this series of researches.
First, the blended (or averaged, or emergent) dynamics is a virtual one and a solution to the blended dynamics may not be generated by any single agent (unless the agents \eqref{eq:eachdyn} are identical), so that the behavior of the blended dynamics can be considered as an emergent one.
This observation may be interpreted as a mathematical model of the fact that a unique group behavior can appear even if none of the individuals displays that behavior.
Second, in order to approximate the behavior of \eqref{eq:eachdyn} for an infinite time horizon, various stability properties are imposed on the blended dynamics.
We do emphasize that each agent need not be stable, as long as their combination, i.e., the blended dynamics, is stable.
This shows that, in a typical situation where there are many stable agents in a group, a few unstable (or malfunctioning, or even malicious) agents may coexist without perturbing the stability of the group.
Again, this approach may explain how the stability of each agent is traded off among the connected agents, or even explain a way to maintain the public good against malicious agents by a majority of good neighbors.

Our study of the limiting behavior when $k$ tends to infinity naturally leads to an application to the synchronization (or, consensus) problem of multi-agent systems.
During the last decade, synchronization of multi-agent systems has been actively studied because of numerous applications in diverse areas, e.g., biology, physics, and engineering.
An initial study of synchronization was about the identical multi-agents \citep{olfati2004consensus,moreau2004stability,Ren05,seo2009consensus}, but the interest soon shifted to the heterogeneous case because uncertainty, disturbance, and noise are prevalent in practice, and so the assumption of identical multi-agents may be too ideal.
Therefore, it may be a natural follow-up to study synchronization for a heterogeneous multi-agent system.
Earlier results in this direction such as \citep{wieland2011internal,kim2011output} have found that each agent must contain a common internal model that is the same across the heterogeneous agents, for synchronization.
If this was not the case, some engineering problems were solved by embedding a common internal model in the consensus controller attached to each heterogeneous agent \citep{kim2011output}.
However, if the multi-agent is not an engineering system such as a network of biological systems that have no common internal model, then achieving synchronization is impossible, and approximate synchronization is studied as an alternative.
This alternative is often achieved with the help of strong coupling (or, a large gain $k$).
To be precise, let us define the term ``(semi-global) practical synchronization.''

\begin{defn1}\label{def:1}
The system \eqref{eq:eachdyn} is said to achieve {\em semi-global practical output synchronization with $\{O_i \in \R^{n \times n}: i \in \NN\}$} if, for every compact set $K \subset \mathbb{R}^{nN}$ and for every $\eta > 0$, there exist $k^*$ such that, for each $k > k^*$ and ${\rm col}(x_1(0), \dots, x_N(0)) \in K$, the solution $\{x_1(t,k),\dots,x_N(t,k)\}$ exists for all $t \ge 0$ and satisfies
\begin{align*}
\limsup_{t \to \infty} \|O_ix_i(t,k) - O_jx_j(t,k)\| \le \eta, \quad \forall i, j \in \mathcal{N}.
\end{align*}
When all $O_i = I_n$, the system is said to achieve {\em semi-global practical state synchronization}.
If the above inequality holds with $\eta = 0$, we remove ``practical,'' and if $K = \R^{nN}$ then we remove ``semi-'' from those terms.
\end{defn1}

Based on this definition, studying the limiting solution offers a simple way to find conditions for the agents to achieve (semi-global) practical output synchronization.
It will be shown that, if the system \eqref{eq:eachdyn} admits the limiting solution with the convergence in \eqref{eq:pwconv} being uniform in $t$, and if the limiting solutions satisfy $\lim_{t \to \infty}\|O_i\xi_i(t)-O_j\xi_j(t)\| = 0$, $\forall i, j \in \NN$, then (semi-global) practical output synchronization with $\{O_i\}$ can be achieved.

Synchronization achieved in this way is different from the so-called ``average consensus'' studied in, e.g., \citep{olfati2004consensus,moreau2004stability,Ren05,Cao08,tuna2009conditions,scardovi2009synchronization,seo2009consensus}, in the sense that the synchronized trajectory does not depend on the initial conditions of the multi-agent system.
Indeed, from the stability imposed on the blended dynamics, it will be seen that the effect of those initial conditions on the limiting solution $\xi_i$ diminishes as time goes on.
This point will be clarified by an example in Section \ref{sec:application0} where we emphasize why forgetting the initial conditions is useful for the so-called plug-and-play operation, which means, for example, that agents can join or leave the network on-line.
Another benefit of the synchronization achieved in this paper over the average consensus is that it is {\em robust} against external disturbance, noise, and/or uncertainty in the agent dynamics.
We do not discuss this point in this paper, but interested readers are referred to \citep{kim2016robustness}.

This paper is organized as follows.
A coordinate change is proposed in Section \ref{sec:main}, with which the system \eqref{eq:eachdyn} is converted into the standard singular perturbation form.
This form enables a formal definition of the blended dynamics to be proposed in Section \ref{sec:estimation}, in which we also demonstrate approximation of the behavior of the system \eqref{eq:eachdyn} under various stability properties imposed on the blended dynamics.
For the synthesis of a multi-agent system on the basis of the analysis performed in Section \ref{sec:estimation}, several special cases are studied in Section \ref{sec:cases}.
Section \ref{sec:application} is devoted to demonstrating the utility of the tool developed in this paper.
In particular, we discuss distributed state estimation in Section \ref{sec:application1}, robust synchronization of heterogeneous Van der Pol oscillators in Section \ref{sec:application2}, distributed estimation of the number of agents in the network in Section \ref{sec:application0}, and a simple distributed optimization in Section \ref{sec:application3}.
All the proofs of the theorems appearing in Section \ref{sec:estimation} are presented in the Appendix.

{\em Notation:}
The Laplacian matrix $\mathcal{L} = [l_{ij}] \in \mathbb{R}^{N \times N}$ of a graph is defined as $\mathcal{L} := \mathcal{D} - \mathcal{A}$, where $\mathcal{A} = [\alpha_{ij}]$ is the adjacency matrix of the graph, and $\mathcal{D}$ is the diagonal matrix whose diagonal entries are determined such that each row sum of $\mathcal{L}$ is zero.
By its construction, it contains at least one eigenvalue of zero, whose corresponding eigenvector is $1_N := [1,\dots,1]^T \in \R^N$, and all the other eigenvalues have nonnegative real parts.
For undirected graphs, the zero eigenvalue is simple if and only if the corresponding graph is connected.
For vectors or matrices $a$ and $b$, ${\rm col}(a,b) := [a^T,b^T]^T$.
For matrices $A_1, \dots, A_k$, we denote by $\text{diag}(A_1, \dots, A_k)$ the block diagonal matrix.
The operation defined by the symbol $\otimes$ is the Kronecker product.
For a set $\Xi \subset \R^n$, $\|x\|_{\Xi}$ denotes the distance between the vector $x \in \R^n$ and $\Xi$; i.e., $\| x \|_{\Xi} := \inf_{y \in \Xi} \| x - y\|$.
The identity matrix of size $m \times m$ is denoted by $I_m$.
In this paper, all positive (semi-)definite matrices are symmetric.
For a matrix $A \in \R^{n \times m}$, ${\rm im}(A) := \{ y \in \R^n : y = A x, x \in \R^m \}$ (we use this notation even when $A$ is a vector in $\R^n$ by treating $A$ as a $n$-by-$1$ matrix), and $\ker(A) := \{ x \in \R^m : Ax = 0 \}$.

\section{Coordinate Change for Singularly Perturbed Form}\label{sec:main}

\subsection{Preliminaries}

The main result is stated under the following assumption.

\begin{assum1}\label{assum:L}
The communication graph induced by the adjacency element $\alpha_{ij}$ is undirected and connected, and thus, the Laplacian matrix $\LL$ is symmetric, having one simple eigenvalue of zero.
\end{assum1}

We introduce a linear coordinate change that is fundamental in the proposed analysis.
The goal is to convert the networked multi-agent system \eqref{eq:eachdyn} with a large coupling gain $k$ into a standard form of singular perturbation analysis \citep{maghenem2016singular}.
Then, we will demonstrate that the behavior of each agent can be approximated on the basis of various stability properties imposed on the blended dynamics to be defined.
It will turn out that the blended dynamics is nothing but the slow dynamics of the singularly perturbed system, i.e., the quasi-steady-state subsystem.

The proposed change of coordinates is composed of a few matrices that can always be found as follows:
\begin{enumerate}
	
	\item For each positive semi-definite matrix $B_i$ for \eqref{eq:eachdyna} (or, $C_i$ for \eqref{eq:eachdynb}), find $W_i \in \mathbb{R}^{n \times p_i}$, $Z_i \in \mathbb{R}^{n \times (n - p_i)}$, and a positive definite matrix $\Lambda_i \in \mathbb{R}^{p_i \times p_i}$, where $p_i$ is the rank of $B_i$ (or, $C_i$), such that $[W_i \; Z_i]$ is an orthogonal matrix and
	\begin{subequations}
	\begin{align}
	B_i &= \begin{bmatrix} W_i & Z_i \end{bmatrix} \begin{bmatrix} \Lambda_i^2 & 0 \\ 0 & 0 \end{bmatrix}\begin{bmatrix} W_i^T \\ Z_i^T \end{bmatrix} \label{eq:Bi} \\
	\text{or,} \quad C_i &= \begin{bmatrix} W_i & Z_i \end{bmatrix} \begin{bmatrix} \Lambda_i^2 & 0 \\ 0 & 0 \end{bmatrix}\begin{bmatrix} W_i^T \\ Z_i^T \end{bmatrix} .  \label{eq:Ci}
	\end{align}
	\end{subequations}
	For future use, let $\oo{W} := {\rm diag}(W_1, \dots, W_N)$, $\oo{Z} := {\rm diag}(Z_1, \dots, Z_N)$, and $\oo{\Lambda} := {\rm diag}(\Lambda_1, \dots, \Lambda_N)$.
	
	\item Find $V_i \in \R^{p_i \times p_o}$ such that, with $\bar p := \sum_{i=1}^N p_i$ and $V := {\rm col}(V_1, \dots, V_N) \in \R^{\bar p \times p_o}$, the columns of $V$ are orthonormal vectors satisfying
	\begin{equation}\label{eq:TInull}
	(D \otimes I_n) \oo{W} \oo{\Lambda} V = 0_{n(N-1) \times p_o}
	\end{equation}
	where $D \in \R^{(N-1) \times N}$ is any matrix satisfying $\ker (D) = {\rm im} (1_N)$,\footnote{An example is
		$$D = \begin{bmatrix} 1 & -1 & & & &  \\ & 1 & -1 & & & \\ & & & \ddots & &  \\ & & & 1 & -1 \end{bmatrix} \in \R^{(N-1) \times N}.$$
	It can be shown that the particular choice of $D$ does not affect the selection of $V$ as long as $\ker (D) = {\rm im} (1_N)$.}
	and $p_o$ is the dimension of ${\rm ker}(D \otimes I_n) \oo{W} \oo{\Lambda}$.

	\item Find $\overline V \in \R^{\bar p \times (\bar p-p_o)}$ such that $[V \; \overline V] \in \R^{\bar p \times \bar p}$ is an orthogonal matrix.

\end{enumerate}

\begin{prop1}\label{prop:p}
\begin{enumerate}
\item [(i)] $p_o \le \min\{p_1,\dots,p_N\} \le n$. 
\item [(ii)] All $W_i \Lambda_i V_i$, $i=1,\cdots,N$, are the same matrix, so let us denote it by $M \in \R^{n \times p_o}$, and the rank of $M$ is $p_o$. 
\item [(iii)] Define
\begin{equation}\label{eq:Q}
Q := \overline V^T \oo{\Lambda} \oo{W}^T (\LL \otimes I_n) \oo{W} \oo{\Lambda} \overline V \quad \in \R^{(\bar p-p_o) \times (\bar p-p_o)}.
\end{equation}
Then, $Q$ is positive definite.
\end{enumerate}
\end{prop1}

\begin{pf}
(i) Since $\ker( D \otimes I_n) = {\rm im} (1_N \otimes I_n)$ by construction, it is seen that
\begin{align*}
&\ker (D \otimes I_n) \oo{W} \oo{\Lambda} = \{ \nu \in \R^{\bar p} : \oo{W} \oo{\Lambda}\nu \in {\rm im} (1_N \otimes I_n) \} \\
&= \{ {\rm col}(\nu_1,\dots,\nu_N) \in \R^{\bar p} : W_1 \Lambda_1 \nu_1 = \cdots = W_N \Lambda_N \nu_N \} \\
&= \{ \nu \in \R^{\bar p} : \oo{W}\oo{\Lambda}\nu = 1_N \otimes c, c \in \cap_{i=1}^N {\rm im} (W_i\Lambda_i) \subset \R^n \}.
\end{align*} 
Since $\nu$ is uniquely determined for each $c \in \cap_{i=1}^N {\rm im}( W_i\Lambda_i) = \cap_{i=1}^N {\rm im}(W_i)$, we have $p_o = \dim \ker (D \otimes I_n) \oo{W}\oo{\Lambda} = \dim \cap_{i=1}^N {\rm im} (W_i)$.
Therefore, it follows that $p_o \le \min \{p_1,\dots,p_N\}$ since $p_i = \dim {\rm im} (W_i)$, and (i) holds since $p_i \le n$.

(ii) By the above construction, we have ${\rm im}  (\oo{W} \oo{\Lambda} V ) \subset \ker (D \otimes I_n )= {\rm im}(1_N \otimes I_n)$.
This implies $W_1 \Lambda_1 V_1 = \dots = W_N \Lambda_N V_N = M$, and thus, we have $V^T\oo{\Lambda}^T\oo{W}^T\oo{W}\oo{\Lambda}V = (\oo{\Lambda}V)^T\oo{\Lambda}V = NM^TM$.
By noting that $\oo{\Lambda}V$ has a full column rank of $p_o$, it is seen that $M^TM \in \R^{p_o \times p_o}$ is nonsingular, which proves (ii).

(iii) It is easily seen that $Q$ in \eqref{eq:Q} is positive semi-definite since $\LL\otimes I_n$ is positive semi-definite.
Hence, it remains to show that $\zeta^T Q \zeta = 0$ for a vector $\zeta \in \R^{\bar p - p_o}$ implies $\zeta = 0$.
Indeed, if $\zeta^T Q \zeta = 0$, then it can be shown that $(\LL \otimes I_n) \oo{W}\oo{\Lambda} \overline V \zeta = 0$.
Noting that $\ker( \LL \otimes I_n)= {\rm im}(1_N \otimes I_n)$ by Assumption \ref{assum:L}, we then have that $(D \otimes I_n) \oo{W}\oo{\Lambda} \overline V \zeta = 0$.
Recalling that $V$ is a basis of $\ker(D \otimes I_n) \oo{W}\oo{\Lambda}$, this implies that $\overline V \zeta \in {\rm im} (V)$.
However, $V$ and $\overline V$ are mutually orthogonal, and thus, $\zeta = 0$.\QEDB
\end{pf}

\subsection{Coordinate Change for \eqref{eq:eachdyna}}

With $x := {\rm col}(x_1, \dots, x_N)$, the system \eqref{eq:eachdyna} is written as
\begin{align}\label{eq:sysa}
\begin{split}
\dot x &= \begin{bmatrix} f_1(t,x_1) \\ \vdots \\ f_N(t,x_N) \end{bmatrix} - k \begin{bmatrix} B_1 & & \\ & \ddots & \\ & & B_N \end{bmatrix} (\LL \otimes I_n) x \\
&=: F(t,x) - k \oo{B} (\LL \otimes I_n) x.
\end{split}
\end{align}
For this system, we propose a coordinate change as
\begin{equation}\label{eq:Pa}
\begin{bmatrix} z \\ z_o \\ w \end{bmatrix} = \begin{bmatrix} \oo{Z}^T \\ V^T \oo{\Lambda}^{-1} \oo{W}^T \\ Q^{-1} \overline V^T \oo{\Lambda} \oo{W}^T (\LL \otimes I_n) \end{bmatrix} x =: P_a x
\end{equation}
where $z \in \R^{nN - \bar p}$, $z_o \in \R^{p_o}$, and $w \in \R^{\bar p - p_o}$.
From this, it is seen that $z = {\rm col}(z_1, z_2, \cdots, z_N)$, where
\begin{equation}\label{eq:z_i_1}
z_i = Z_i^T x_i \quad \in \R^{n-p_i}, \qquad i \in \NN.
\end{equation}
Thus, the state $z_i$ can be considered as a projected component of $x_i$ on ${\rm im} (Z_i)$.
It is also noted that $z_o$ is a weighted sum of $x_i$'s:
\begin{equation}
\label{eq:z_o}
z_o = \sum_{i=1}^N V_i^T \Lambda_i^{-1} W_i^T x_i.
\end{equation}
We now claim that $P_a^{-1}$ is given by
\begin{equation}
\label{eq:Painv}
P_a^{-1} = \begin{bmatrix} \oo{Z} - \oo{W} \oo{\Lambda} L, & \oo{W} \oo{\Lambda} V, & \oo{W} \oo{\Lambda} \overline V \end{bmatrix}
\end{equation}
where $L \in \R^{\bar p \times (nN - \bar p)}$ is defined as
\begin{equation}
\label{eq:L}
L = \begin{bmatrix} L_1 \\ \vdots \\ L_N \end{bmatrix} := \overline V Q^{-1} \overline V^T \oo{\Lambda} \oo{W}^T (\LL \otimes I_n) \oo{Z}
\end{equation}
with $L_i \in \R^{p_i \times (nN - \bar p)}$.
This claim can be proved by verifying $P_aP_a^{-1}=I_{nN}$.
For this, one may use the definitions of $Q$ and $L$ with the facts that $[V, \overline V]$ and $[\oo{W}, \oo{Z}]$ are orthogonal matrices, and $(\LL \otimes I_n) \oo{W}\oo{\Lambda} V = 0$.
Similarly, the following claim can also be proved:
\begin{align}
&P_a\oo{B}(\LL\otimes I_n)P_a^{-1} = \begin{bmatrix} \oo{Z}^T \\ V^T \oo{\Lambda}^{-1} \oo{W}^T \\ Q^{-1} \overline V^T \oo{\Lambda} \oo{W}^T (\LL \otimes I_n) \end{bmatrix} \label{eq:ct} \\
&\times \oo{B} (\LL \otimes I_n) \begin{bmatrix} \oo{Z}-\oo{W}\oo{\Lambda} L, & \oo{W}\oo{\Lambda} V, & \oo{W}\oo{\Lambda} \overline V \end{bmatrix} = \begin{bmatrix} 0 & 0 & 0 \\ 0 & 0 & 0 \\ 0 & 0 & Q \end{bmatrix}. \notag
\end{align}
To do this, one may also use the facts that $\oo{B} = \oo{W}\oo{\Lambda}^2\oo{W}^T$ and $\oo{\Lambda}^2 = \oo{\Lambda} (VV^T + \overline V \overline V^T) \oo{\Lambda}$.
With \eqref{eq:Pa} and $P_a^{-1}$ of \eqref{eq:Painv}, we have that, for all $i \in \NN$,
\begin{equation}\label{eq:inv}
x_i = Z_i z_i - W_i\Lambda_iL_i z + W_i\Lambda_iV_i z_o + W_i\Lambda_i\overline V_i w
\end{equation}
where $\overline V = {\rm col}(\overline V_1, \ldots, \overline V_N)$ with $\overline V_i \in \R^{p_i \times (\bar p - p_o)}$.
Taking time derivatives of \eqref{eq:Pa} with \eqref{eq:ct} and \eqref{eq:inv} in mind yields the following representation of the system \eqref{eq:sysa}:
	\begin{align}\label{eq:tsysa}
	\dot z_i &= Z_i^T f_i(t, Z_i z_i - W_i\Lambda_i(L_i z - V_i z_o - \overline V_i w)) \nonumber \\
	\dot z_o &= \sum_{i=1}^N V_i^T\Lambda_i^{-1}W_i^T \notag \\
	& \qquad \times f_i(t, Z_i z_i - W_i\Lambda_i(L_i z - V_i z_o - \overline V_i w)) \nonumber \\
	\epsilon \dot w &= -Q w + \epsilon \Theta^T F(t, \oo{Z} z - \oo{W}\oo{\Lambda}(L z - V z_o - \overline V w))
	\end{align}
for $i \in \NN$ where $\epsilon := 1/k$ and $\Theta^T := Q^{-1} \overline V^T \oo{\Lambda} \oo{W}^T (\LL \otimes I_n)$ for convenience.
With $Q$ being positive definite and $k$ sufficiently large, it is clear that this system is in the standard singular perturbation form.

\subsection{Coordinate Change for \eqref{eq:eachdynb}}

Regarding the system \eqref{eq:eachdynb}, one can show that \eqref{eq:eachdynb} can be compactly written as
\begin{align}\label{eq:sysb}
\begin{split}
\dot x &= \begin{bmatrix} f_1(t,x_1) \\ \vdots \\ f_N(t,x_N) \end{bmatrix} - k (\LL \otimes I_n) \begin{bmatrix} C_1 & & \\ & \ddots & \\ & & C_N \end{bmatrix} x \\
&=: F(t,x) - k (\LL \otimes I_n) \oo{C} x.
\end{split}
\end{align}
This time, we propose a coordinate change as
\begin{equation}\label{eq:Pb}
\begin{bmatrix} z \\ z_o \\ w \end{bmatrix} = \begin{bmatrix} \oo{Z}^T - L^T \oo{\Lambda} \oo{W}^T \\ V^T \oo{\Lambda} \oo{W}^T \\ \overline V^T \oo{\Lambda} \oo{W}^T \end{bmatrix} x =: P_b x
\end{equation}
where $z \in \R^{nN - \bar p}$, $z_o \in \R^{p_o}$, and $w \in \R^{\bar p - p_o}$, and $L$ is defined in \eqref{eq:L}.
Then it follows immediately from \eqref{eq:Painv} that $P_b = P_a^{-T}$.
From this, the state $z_i$ can be written as
\begin{equation}
z_i = Z_i^T x_i - \sum_{j=1}^N L_{ji}^T \Lambda_j W_j^T x_j
\end{equation}
where $L_{ji} \in \R^{p_j \times (n-p_i)}$ is such that $L_j = [L_{j1}, \dots, L_{jN}]$.
This looks more complicated compared to \eqref{eq:z_i_1}.
Instead, we have a simpler version of \eqref{eq:z_o}:
\begin{equation}
\label{eq:z_ob}
z_o = M^T \sum_{i=1}^N x_i
\end{equation}
because $W_i \Lambda_i V_i = M$ for all $i \in \NN$.
The inverse $P_b^{-1}$ can also be easily obtained by $P_a^T$:
\begin{equation}
\label{eq:Pbinv}
P_b^{-1} = \begin{bmatrix} \oo{Z}, & \oo{W} \oo{\Lambda}^{-1} V, & (\LL \otimes I_n) \oo{W} \oo{\Lambda} \overline V Q^{-1} \end{bmatrix}.
\end{equation}
This leads to
\begin{equation}\label{eq:invb}
x_i = Z_i z_i + W_i\Lambda_i^{-1}V_i z_o + \Theta_i w
\end{equation}
where $\Theta = {\rm col}(\Theta_1,\dots,\Theta_N) = (\LL \otimes I_n) \oo{W} \oo{\Lambda} \overline V Q^{-1}$ with $\Theta_i = \sum_{j \in \NN_i} \alpha_{ij} (W_i \Lambda_i \overline V_i - W_j \Lambda_j \overline V_j) Q^{-1} \in \R^{n \times (\bar p - p_o)}$.
Now, because $\oo{C} = \oo{W} \oo{\Lambda}^2 \oo{W}^T$, one can recycle the relation \eqref{eq:ct} by taking the transpose to obtain
\begin{equation}
\label{eq:ctb}
P_b (\LL \otimes I_n) \oo{C} P_b^{-1} = \begin{bmatrix} 0 & 0 & 0 \\ 0 & 0 & 0 \\ 0 & 0 & Q \end{bmatrix}.
\end{equation}
Finally, the time derivatives of \eqref{eq:Pb} yield the following representation of the system \eqref{eq:sysb}:
	\begin{align}\label{eq:tsysb}
	\dot z_i &= Z_i^T f_i(t, Z_i z_i + W_i\Lambda_i^{-1}V_i z_o + \Theta_i w) \nonumber \\
	& \hspace{-3mm} - \sum_{j=1}^N L_{ji}^T \Lambda_j W_j^T f_j(t, Z_j z_j + W_j\Lambda_j^{-1}V_j z_o + \Theta_j w) \notag \\
	\dot z_o &= M^T \sum_{i=1}^N f_i(t, Z_i z_i + W_i\Lambda_i^{-1}V_i z_o + \Theta_i w)
	\nonumber \\
	\epsilon \dot w &= -Q w + \epsilon \overline V^T \oo{\Lambda} \oo{W}^T F(t, \oo{Z} z + \oo{W} \oo{\Lambda}^{-1} V z_o + \Theta w) 
	\end{align}
for $i \in \NN$ where $\epsilon = 1/k$.
It is clear that this system is again in the standard singular perturbation form.

\section{Behavior of the Multi-agent System}\label{sec:estimation}

The system \eqref{eq:eachdyn} has now been transformed to the standard singular perturbation form, from which it can be seen that the boundary-layer subsystem is $dw/d\tau = -Q w$, and the quasi-steady-state system is obtained with $w \equiv 0$.
We call this quasi-steady-state system the {\em blended dynamics} for \eqref{eq:eachdyn}.
It is seen from \eqref{eq:tsysa} that the blended dynamics for \eqref{eq:eachdyna} is given by
\begin{subequations}\label{eq:qss}
\begin{align}\label{eq:qssa}
\begin{split}
\dot {\hat z}_i &= Z_i^T f_i(t, Z_i \hat z_i - W_i \Lambda_i L_i \hat z + M \hat z_o), \quad i \in \NN, \\
\dot {\hat z}_o &= \sum_{i=1}^N V_i^T \Lambda_i^{-1} W_i^T f_i(t, Z_i \hat z_i - W_i \Lambda_i L_i \hat z + M \hat z_o),  \\ 
\hat z_i&(0) = Z_i^T x_i(0), \quad \hat z_o(0) = \sum_{i=1}^N V_i^T \Lambda_i^{-1} W_i^T x_i(0) 
\end{split}
\end{align}
for $i \in \NN$, where $\hat z_i \in \R^{n-p_i}$, $\hat z = {\rm col}(\hat z_1, \dots, \hat z_N)$, and $\hat z_o \in \R^{p_o}$, so that it is a dynamic system of dimension $m := nN-(\bar p - p_o)$.
Similarly, from \eqref{eq:tsysb}, the blended dynamics for \eqref{eq:eachdynb} is given by
\begin{align}\label{eq:qssb}
\begin{split}
\dot {\hat z}_i &= Z_i^T f_i(t, Z_i \hat z_i + W_i \Lambda_i^{-1} V_i \hat z_o) \\
&- \sum_{j=1}^N L_{ji}^T\Lambda_j W_j^T f_j(t, Z_j \hat z_j + W_j \Lambda_j^{-1} V_j \hat z_o), \quad i \in \NN, \\
\dot {\hat z}_o &= M^T \sum_{i=1}^N f_i(t, Z_i \hat z_i + W_i \Lambda_i^{-1} V_i \hat z_o), \\
\hat z_i(0) &= Z_i^T x_i(0) - \sum_{j=1}^N L_{ji}^T\Lambda_j W_j^T x_j(0), \\
\hat z_o(0) &= M^T \sum_{i=1}^N x_i(0) . 
\end{split}
\end{align}
\end{subequations}

Now, we can approximate the actual trajectory $x_i$ for all $i \in \NN$ when $k$ tends to infinity, as long as the solution to the blended dynamics exists for all future time.

\begin{prop1}\label{prop:extreme}
	Under Assumption \ref{assum:L}, assume that, for given initial conditions $x_i(0)$, $i \in \NN$ of \eqref{eq:eachdyn}, the solutions $\hat z_i(t)$ and $\hat z_o(t)$ to \eqref{eq:qss} exist for all $t \ge 0$.
	Then, the system \eqref{eq:eachdyn} admits the limiting solution; that is, for each $t > 0$, 
	\begin{subequations}
	\begin{equation}
	\lim_{k \to \infty} x_i(t,k) = Z_i \hat z_i(t) - W_i\Lambda_iL_i \hat z(t) + M \hat z_o(t) =: \xi_i(t) \label{eq:prop1a}	
	\end{equation}
	or,
	\begin{equation}
	\lim_{k \to \infty} x_i(t,k) = Z_i \hat z_i(t) + W_i\Lambda_i^{-1}V_i \hat z_o(t) =: \xi_i(t) \label{eq:prop1b}
	\end{equation}
	\end{subequations}
	for all $i \in \NN$, where $x_i(\cdot,k)$ is the solution of \eqref{eq:eachdyn} with the coupling gain $k$.
\end{prop1}

\begin{rem1}
    Note that we define $\xi_i(t)$ for $t > 0$ but not at $t=0$ because $x_i(0,k)=x_i(0)$ for all $k$ and $\lim_{k\to\infty}x_i(0,k)$ need not be the same as $\lim_{t \to 0^+} \xi_i(t)$.
\end{rem1}

\begin{rem1}
Note that the choice of $V$ does not alter the limiting solution $\xi_i$.
This is because, given one choice of $V$, any other choices $\tilde{V}$ can be represented as
$$\tilde{V} = V \mathsf{O},$$
where $\mathsf{O} \in \mathbb{R}^{p_o \times p_o}$ is an orthogonal matrix.
Then, this only alters $z_o$ to $\tilde{z}_o$, where $\tilde{z}_o = \mathsf{O}^Tz_o$.
Therefore, the limiting solution $\xi_i$ is invariant:
\begin{subequations}
\begin{align}
\begin{split}
\xi_i(t) &= Z_i\hat{z}_i(t) - W_i\Lambda_iL_i\hat{z}(t) + W_i\Lambda_iV_i\hat{z}_o(t) \\
&= Z_i\hat{z}_i(t) - W_i\Lambda_iL_i\hat{z}(t) + W_i\Lambda_i\tilde{V}_i\mathsf{O}^T\hat{z}_o(t)
\end{split}
\end{align}
or,
\begin{align}
\begin{split}
\xi_i(t) &= Z_i\hat{z}_i(t) + W_i\Lambda_i^{-1}V_i\hat{z}_o(t) \\
&= Z_i\hat{z}_i(t) + W_i\Lambda_i^{-1}\tilde{V}_i\mathsf{O}^T\hat{z}_o(t).
\end{split}
\end{align}
\end{subequations}
The same reasoning holds for the blended dynamics, and thus, the stability assumptions for the blended dynamics given in the rest of this paper are independent of the choice of $V$.
\end{rem1}

While the convergence in Proposition \ref{prop:extreme} is point-wise in time $t$, it will be shown that this convergence becomes uniform in time if the blended dynamics is stable in a certain sense.
The forthcoming three theorems specify these stabilities precisely and show how the solution of the multi-agent system \eqref{eq:eachdyn} behaves based on three different stability assumptions.
Here, it is emphasized again that we require stability for the blended dynamics but not for the individual agents.

\begin{thm1}\label{thm:contraction}
Under Assumption \ref{assum:L}, assume that the blended dynamics \eqref{eq:qss} is contractive.\footnote{We say that $\dot x = f(t, x)$ is contractive if there exist a positive definite matrix $H_c$ and a positive constant $\lambda_c$ such that $H_c(\partial f/\partial x)(t, x) + (\partial f/\partial x)^T(t, x)H_c \le -\lambda_cH_c$, $\forall x \in \mathbb{R}^m$, $\forall t \ge 0$ \citep{pavlov2005convergent}.}
Then, for any compact set $K \subset \mathbb{R}^{nN}$ and for any $\eta > 0$, there exists $k^* > 0$ such that, for each $k > k^*$ and $x(0) \in K$, the solution $x(t, k)$ to \eqref{eq:eachdyn} exists for all $t \ge 0$, and satisfies
$$\limsup_{t \to \infty} \|x_i(t, k) - \xi_i(t)\| \le \eta, \quad \forall i \in \mathcal{N}.$$
\end{thm1}

Theorem \ref{thm:contraction} does not require the system \eqref{eq:qss} to have an equilibrium point.
However, it relies on the contraction property of the blended dynamics.
The following theorem is for the case when there is a compact attractor, e.g., an asymptotically stable equilibrium, of the blended dynamics.

\begin{thm1}\label{thm:ps}
Under Assumption \ref{assum:L}, assume that there is a nonempty compact set $\AAA_z \subset \R^{m}$ that is uniformly asymptotically stable for the blended dynamics \eqref{eq:qss}.
Let $\DDD_z \supset \AAA_z$ be an open set contained in the domain of attraction of $\AAA_z$, and let
\begin{subequations}
\begin{multline}
\DDD_x :=  \{ (\oo{Z}-\oo{W} \oo{\Lambda} L) \hat{z} + (1_N \otimes M) \hat z_o + \oo{W} \oo{\Lambda} \overline{V} w \in \R^{nN} \\ 
: {\rm col}(\hat{z}, \hat z_o) \in \DDD_z, w \in \mathbb{R}^{\bar{p}-p_o} \}
\end{multline}
\begin{multline}
\text{or, } \DDD_x :=  \{ \oo{Z} \hat{z} + \oo{W}\oo{\Lambda}^{-1}V \hat z_o + \Theta w \in \R^{nN} \\
: {\rm col}(\hat{z}, \hat z_o) \in \DDD_z, w \in \mathbb{R}^{\bar{p}-p_o} \}.
\end{multline}
\end{subequations}
Then, for any compact set $K \subset \DDD_x \subset \R^{nN}$ and for any $\eta >0$ and $\tau > 0$, there exists $k^* > 0$ such that for each $k > k^*$ and $x(0) \in K$, the solution $x(t, k)$ to \eqref{eq:eachdyn} exists for all $t \ge 0$, and satisfies 
\begin{align}
\|x_i(t, k) - \xi_i(t) - \tilde{\xi}_i(kt) \| &\le \eta, \;\; \forall t \in [0, \tau], i \in \NN \label{eq:setconv} \\
\limsup_{t \to \infty} \|x(t, k)\|_{\AAA_x} &\le \eta,  \label{eq:setconv2}
\end{align}
where $\tilde \xi_i$ is an exponentially decaying function such that $\lim_{t \to \infty} \tilde \xi_i(t) = 0$, and 
\begin{subequations}\label{eq:setA}
\begin{align}
&\AAA_x := \{(\oo{Z}-\oo{W}\oo{\Lambda} L)\hat{z} + (1_N \otimes M) \hat z_o : {\rm col}(\hat{z}, \hat z_o) \in \AAA_z \} \\
&\text{or,}\;\; \AAA_x := \{\oo{Z} \hat{z} + \oo{W}\oo{\Lambda}^{-1}V \hat z_o : {\rm col}(\hat{z}, \hat z_o) \in \AAA_z \}.
\end{align}
\end{subequations}
\end{thm1}

If a vanishing condition is appended to the multi-agent system, we obtain a stronger result on top of Theorem \ref{thm:ps}.

\begin{thm1}\label{thm:as}
In addition to the assumptions of Theorem \ref{thm:ps}, assume that
\begin{enumerate}
\item for all $x \in \AAA_x$ and $t \ge 0$, 
\begin{subequations}\label{eq:invariance}
\begin{gather}
\Theta^T F(t,x) = 0 \\
\text{or, } \;\; \overline{V}^T \oo{\Lambda} \oo{W}^T F(t,x) = 0, 
\end{gather}	
\end{subequations}
\item the set $\AAA_z$ is locally exponentially stable for the blended dynamics \eqref{eq:qss}.
\end{enumerate}
Then, for any compact set $K \subset \DDD_x$, there exists $k^{*} > 0$ such that for each $k > k^{*}$ and $x(0) \in K$, the solution $x(t, k)$ to \eqref{eq:eachdyn} exists for all $t \ge 0$, and satisfies
	\begin{equation}\label{eq:convergence}
		\lim_{t \to \infty} \| x(t, k) \|_{\AAA_x} = 0.
	\end{equation}
\end{thm1}

\begin{rem1}\label{rem:cond}
Recall that the objective in Theorem \ref{thm:as} is to achieve asymptotic convergence to the set $\AAA_x$.
For this purpose, it is necessary that $\AAA_x$ is invariant under \eqref{eq:eachdyn}.
By the structure of the set $\AAA_x$ given in \eqref{eq:setA}, the state $w$ should be zero when $x$ belongs to $\AAA_x$.
According to \eqref{eq:tsysa} or \eqref{eq:tsysb}, this in turn requires \eqref{eq:invariance} to hold for all $x \in \AAA_x$ and $t \ge 0$.
\end{rem1}

\begin{rem1}\label{rem:global}
If the multi-agent system \eqref{eq:eachdyn} is affine such as $f_i(t,x_i) = A_i x_i + g_i(t)$ with a constant matrix $A_i$ and a bounded external input $g_i(t)$, then the results of the above theorems become global, because global behavior is the same as local behavior for linear systems.
\end{rem1}

For the multi-agent system \eqref{eq:eachdynb}, the conclusion of Theorem \ref{thm:contraction} implies practical output synchronization with $\{O_i = C_i, i \in \NN \}$.
This follows from the observation that $C_i\xi_i = W_i\Lambda_i^2 W_i^T (Z_i \hat z_i + W_i \Lambda_i^{-1} V_i \hat z_o) = M \hat z_o = C_j\xi_j$.
On the other hand, if the conditions of Theorem \ref{thm:ps} are met for \eqref{eq:eachdynb}, then a large $k$ can make $\limsup_{t\to\infty}\|x(t, k)\|_{\AAA_x}$ arbitrarily small.
This implies that practical output synchronization is achieved with $O_i = C_i$ because $\AAA_x$ is contained in an output synchronization manifold defined as $\{ {\rm col}(x_1, \dots, x_N) \in \mathbb{R}^{nN} : C_ix_i = C_jx_j, \forall i \neq j \}$ since $\oo{C} x = \oo{W}\oo{\Lambda}^2\oo{W}^Tx = (1_N \otimes M)\hat z_o$ for any $x \in \AAA_x$.
If the assumptions of Theorem \ref{thm:as} hold for \eqref{eq:eachdynb}, then output synchronization is achieved.

It is noted that the blended dynamics \eqref{eq:qss} implicitly contains the information of the network graph through the matrix $L_i$.
Therefore, in the cases when the information on the graph structure is not available, stability verification of the blended dynamics, which is necessary for applying three theorems in this section, becomes difficult.
In the next section, we present special cases where the blended dynamics does not depend on the graph structure, which may be found useful for applications.

\section{Special Cases}\label{sec:cases}

\subsection{Identical Coupling Matrices}\label{sec:case1}

When all the $B_i$'s in \eqref{eq:eachdyna} are identical to a positive semi-definite matrix $B_o$,\footnote{If all the $C_i$'s in \eqref{eq:eachdynb} are identical to, say, $C_o$, then this case can also be considered as the case that all $B_i$'s are the same as $B_o=C_o$, because the matrix $C_i$ can go in front of the summation in \eqref{eq:eachdynb}.} we have, referring to \eqref{eq:Bi}, that $p_1=\cdots=p_N =: p_o$, and that $W_1 = \cdots = W_N =: W_o \in \R^{n \times p_o}$, $Z_1 = \cdots = Z_N =: Z_o \in \R^{n \times (n-p_o)}$, and $\Lambda_1 = \cdots = \Lambda_N =: \Lambda_o \in \R^{p_o \times p_o}$.
Then, since $\oo{W}\oo{\Lambda} = I_N \otimes W_o \Lambda_o$ in this case so that $(D \otimes I_n) \oo{W} \oo{\Lambda} = D \otimes W_o \Lambda_o$, we can take any square matrix $V_o \in \R^{p_o \times p_o}$ such that the columns of $V = {\rm col}(V_o,\dots,V_o) \in \R^{Np_o \times p_o}$ are an orthonormal basis of $\ker (D \otimes W_o \Lambda_o)$.
This implies that $V^T V = \sum_{i=1}^N V_o^T V_o = N V_o^TV_o = I_{p_o}$, from which we obtain that $V_o V_o^T = (1/N)I_{p_o}$.
Finally, since $\oo{W} = I_N \otimes W_o$, $\oo{Z} = I_N \otimes Z_o$, and $W_o^T Z_o = 0$, the matrix $L$ of \eqref{eq:L} becomes a zero matrix.

In this case, we can further simplify the expression of \eqref{eq:qssa} because all $W_i$, $\Lambda_i$, and $V_i$ are the same.
Indeed, with all the above relations and with the new variable defined by
\begin{equation}
\hat s := \Lambda_o V_o \hat z_o \quad \in \R^{p_o},
\end{equation}
the blended dynamics \eqref{eq:qssa} becomes
\begin{align}
\dot {\hat z}_i &= Z_o^T f_i(t, Z_o \hat z_i + W_o \hat s) \quad\quad\quad\,\,\,\, \in \R^{n-p_o}, \quad i \in \NN, \notag \\
\dot {\hat s} &= \frac1N \sum_{i=1}^N W_o^T f_i(t, Z_o \hat z_i + W_o \hat s) \,\, \in \R^{p_o} \label{eq:scase1} \\
\hat z_i&(0) = Z_o^T x_i(0), \quad \hat s(0) = \frac1N \sum_{i=1}^N W_o^T  x_i(0). \notag
\end{align}
It is noted that, with $L=0$, there is no direct interaction among $\hat z_i$-dynamics, while each $\hat z_i$ interacts with $\hat s$-dynamics so that the interactions among the $\hat z_i$'s are indirect through the state variable $\hat s$.
An example of this case will be given in Section \ref{sec:application2}.
In particular, the vector field $f_i$ is split into two parts: $W_o^T f_i$ and $Z_o^T f_i$.
The former is averaged for the $\hat s$-dynamics, while the latter remains, comprising a reduced-order multi-agent system.
This is done regardless of the particular structure of the given network graph, as long as it is connected.

In this case, the limiting solution in Proposition \ref{prop:extreme} is given by
\begin{equation}\label{eq:conv1}
	\xi_i(t) = Z_o \hat z_i(t) + W_o \hat s(t).
\end{equation}
It is seen that each $\xi_i$ is affected by the variable $\hat s$, which is common for all $i$, and by the variable $\hat z_i$ but not by $\hat z_j$ for $j \not = i$.
In fact, $\hat z_i$-dynamics is a subsystem of the agent $i$ itself.
Therefore, if $x_i(t)$ approaches $\xi_i(t)$ for all $i \in \NN$, then the question of whether the $x_i$'s achieve (practical) state synchronization depends on whether $\hat z_i$ achieves (practical) state synchronization as time tends to infinity, since $Z_o$ has full column rank.

\subsection{Non-identical Positive Definite Coupling Matrices}\label{sec:case2}

If all the coupling matrices $B_i$ in \eqref{eq:eachdyna} are positive definite (but not necessarily identical), then $p_1 = \cdots = p_N = p_o = n$.
In this case, we can take $W_i = I_n$ for all $i$ so that $W_i \Lambda_i^2 W_i^T = B_i$ with $\Lambda_i = \sqrt{B_i}$, and the matrices $Z_i$ and $L$, and the state $\hat z_i$ are null for all $i \in \NN$.
From Proposition \ref{prop:p}, the matrix $V_i$ can be obtained as $V_i = \sqrt{B_i^{-1}}M$ for all $i \in \NN$, where for $V = {\rm col}(V_1, \dots, V_N)$ to have orthonormal columns, the matrix $M$ should satisfy 
$$V^T V = \sum_{i=1}^N V_i^T V_i = \sum_{i=1}^N M^T \sqrt{B_i^{-1}} \sqrt{B_i^{-1}} M = I_n.$$
By letting $M = \sqrt{(\sum_{i=1}^N B_i^{-1})^{-1}}$, then, the blended dynamics \eqref{eq:qssa}, with a new variable $\hat s := M \hat z_o$, becomes
\begin{subequations}
\begin{align}\label{eq:nonidpd1}
\begin{split}
\dot {\hat s} &= \left(\sum_{i=1}^N B_i^{-1}\right)^{-1} \sum_{i=1}^N B_i^{-1} f_i(t, \hat s) \\
\hat s(0) &= \left(\sum_{i=1}^N B_i^{-1}\right)^{-1} \sum_{i=1}^N B_i^{-1} x_i(0).
\end{split}
\end{align}

Similarly, if all the coupling matrices $C_i$ in \eqref{eq:eachdynb} are positive definite, then one can find $W_i = I_n$, $\Lambda_i = \sqrt{C_i}$, and $V_i = \sqrt{C_i^{-1}} M$ with $M = \sqrt{(\sum_{i=1}^N C_i^{-1})^{-1}}$.
This leads to the blended dynamics \eqref{eq:qssb}, with a new variable $\hat s := M \hat z_o$:
\begin{align}\label{eq:nonidpd2}
\begin{split}
\dot {\hat s} &= \left(\sum_{i=1}^N C_i^{-1}\right)^{-1} \sum_{i=1}^N f_i(t, C_i^{-1} \hat s) \\
\hat s(0) &= \left(\sum_{i=1}^N C_i^{-1}\right)^{-1} \sum_{i=1}^N x_i(0).
\end{split}
\end{align}
\end{subequations}
Another way to arrive at \eqref{eq:nonidpd2} is the coordinate change $\bar x_i := C_i x_i$ for \eqref{eq:eachdynb}, which yields
\begin{align*}
\dot {\bar x}_i = C_i f_i(t, C_i^{-1} \bar x_i) + k C_i \sum_{j \in \NN_i} \alpha_{ij} (\bar x_j - \bar x_i)
\end{align*}
that resembles \eqref{eq:eachdyna}, and use \eqref{eq:nonidpd1} with $B_i = C_i$.

The limiting solution in this case is given by
\begin{equation}\label{eq:33}
\xi_i(t) = \begin{cases} \hat s(t), & \text{for \eqref{eq:eachdyna},} \\ C_i^{-1} \hat s(t), & \text{for \eqref{eq:eachdynb}.} \end{cases}
\end{equation}
In this special case, practical state synchronization is always achieved for \eqref{eq:eachdyna} as long as the conditions of Theorem \ref{thm:ps} are met, because of \eqref{eq:33}.

\subsection{Identical \& Positive Definite Coupling Matrices}\label{sec:case3}

If all the $B_i$'s are identical to a positive definite matrix $B_o$, then, by letting $W_o = I_n$ for the case of Section \ref{sec:case1}, or $B_i = B_o$ in Section \ref{sec:case2}, we obtain the blended dynamics as
\begin{align}\label{eq:scase2}
\begin{split}
\dot {\hat s} = \frac1N \sum_{i=1}^N f_i(t, \hat s), \quad \hat s(0) = \frac1N \sum_{i=1}^N x_i(0) \quad \in \R^n.
\end{split}
\end{align}
In this case, the limiting solution is $\xi_i(t) = \hat s(t)$, from which, we obtain state synchronization by Theorem \ref{thm:as} and practical state synchronization by Theorem \ref{thm:contraction} and Theorem \ref{thm:ps}.
(This means that we recover the results of \citep{kim2012practical,kim2013robustness,kim2016robustness}.)

\section{Applications}\label{sec:application}

\subsection{Distributed State Estimation}\label{sec:application1}

Consider a linear system
$$\dot{\omega} = S\omega \; \in \R^n, \quad \nu = \begin{bmatrix} \nu_1 \\ \vdots \\ \nu_N \end{bmatrix} = \begin{bmatrix} G_1 \\ \vdots \\ G_N \end{bmatrix} \omega = G \omega, \; \nu_i \in \R^{q_i}$$
where $\omega \in \R^n$ is the state to be estimated, and $\nu$ is the measurement output.
It is supposed that there are $N$ distributed nodes, and each node $i$ can access the measurement $\nu_i \in \R^{q_i}$ only (where often $q_i=1$).
We assume that the pair $(G,S)$ is detectable, while each pair $(G_i,S)$ is not necessarily detectable as in \citep{olfati2007distributed, bai2011distributed, kim2016distributed}.
Each node is allowed to communicate its internal state to its neighboring nodes.
The question is how to construct a dynamic system for each node that estimates $\omega(t)$.
See, e.g., \citep{mitra2016approach,kim2016distributedlue} for more details on this distributed state estimation problem.

To solve the problem, we first employ the detectability decomposition for each node, that is, for each pair $(G_i,S)$.
With $p_i$ being the dimension of the undetectable subspace of the pair $(G_i,S)$, let $[Z_i, W_i]$ be an orthogonal matrix, where $Z_i \in \R^{n \times (n-p_i)}$ and $W_i \in \R^{n \times p_i}$, such that
$$\begin{bmatrix} Z_i^T \\ W_i^T \end{bmatrix} S [Z_i \; W_i]
= \begin{bmatrix} \bar{S}_i^{11} & 0 \\ \bar{S}_i^{21} & \bar{S}_i^{22} \end{bmatrix}, \quad 
G_i [Z_i \; W_i] = [\bar{G}_i \; 0]$$
and the pair $(\bar{G}_i, \bar{S}_i^{11})$ is detectable.
Then, pick a matrix $\bar U_i \in \R^{(n-p_i) \times q_i}$ such that $\bar S_i^{11} - \bar U_i \bar G_i$ is Hurwitz, and define $U_i := Z_i \bar U_i \in \R^{n \times q_i}$.

The distributed state observer that we are proposing is
\begin{align}\label{eq:dist_state_obs}
\dot{\hat{\omega}}_i = S \hat{\omega}_i + U_i (\nu_i - G_i\hat{\omega}_i) + k W_iW_i^T \sum_{j=1}^N \alpha_{ij} (\hat \omega_j - \hat \omega_i)
\end{align}
where $k$ is sufficiently large.
Here, the first two terms on the right-hand side look like a typical state observer, but due to the lack of detectability of $(G_i,S)$, it cannot yield stable error dynamics.
Therefore, the diffusive coupling of the third term exchanges the internal state with the neighbors, compensating for the lack of information on the undetectable parts.
Recalling that $W_i^T\omega$ comprises the undetectable part of $\omega$ by $\nu_i$ in the decomposition above, it is noted that the coupling term compensates for only the undetectable portion in the observer.
As a result, the coupling matrix $W_i W_i^T$ is rank-deficient in general.
This point is in sharp contrast to the previous results such as \citep{kim2016distributedlue}, where the coupling term is nonsingular so that the analysis is more complicated.

With $x_i := \hat \omega_i - \omega$ and $B_i = W_i W_i^T$, the error dynamics becomes
$$\dot x_i = (S - U_iG_i)x_i + k B_i \sum_{j=1}^N \alpha_{ij} (x_j-x_i), \qquad i \in \NN.$$
This is precisely the multi-agent system \eqref{eq:eachdyna}, where in this case the matrices $Z_i$ and $W_i$ have implications related to detectable decomposition.
In particular, from the detectability of the pair $(G,S)$, it is seen that $\cap_{i=1}^N {\rm im} (W_i) = \cap_{i=1}^N \ker (Z_i^T) = \{0\}$, by recalling that $\ker (Z_i^T)$ is the undetectable subspace of the pair $(G_i, S)$.
This implies that $p_o=0$ by the construction, $V$ is null, and thus, $\overline{V}$ can be chosen to be the identity matrix.
With them, the blended dynamics is given by, with the state $\hat z_o$ being null,
\begin{align*}
	&\dot{\hat{z}}_i = Z_i^T (S-U_iG_i) (Z_i\hat{z}_i - W_i\Lambda_i L_i \hat{z}) \\
	&= (Z_i^T S - \bar U_i \bar G_i Z_i^T) (Z_i \hat z_i - W_i \Lambda_i L_i \hat z) = (\bar S_i^{11} - \bar U_i \bar G_i) \hat z_i
\end{align*}
for $i \in \NN$, where we have used the fact that $Z_i^T S W_i = 0$ and $Z_i^T W_i = 0$.

Noting that $\bar{S}_i^{11} - \bar{U}_i\bar{G}_i$ is Hurwitz for all $i \in \mathcal{N}$, it is seen that the set $\AAA_z = \{0\} \subset \R^{nN-\bar p}$ is globally exponentially stable.
In addition, the condition (1) of Theorem \ref{thm:as} holds since, in this case, $\AAA_x = \{0\} \subset \R^{nN}$ and $F(t, x)$ is a linear function of $x$ that is zero on $\AAA_x$.
Therefore, Theorem \ref{thm:as} and Remark \ref{rem:global} yields the following.

\begin{cor1}
Suppose that Assumption \ref{assum:L} holds and that $(G, S)$ is detectable.
Then, there exists $k^* > 0$ such that for each $k > k^*$ and $\hat \omega_i(0) \in \mathbb{R}^{n}$, $i \in \NN$, the solution of \eqref{eq:dist_state_obs} exists for all $t \ge 0$, and satisfies
$$\lim_{t \to \infty} \|\hat{\omega}_i(t) - \omega(t)\| = 0, \quad \forall i \in \mathcal{N}.$$
\end{cor1}

\subsection{Synchronization of Heterogeneous Van der Pol Oscillators}\label{sec:application2}

Consider a network of heterogeneous Van der Pol oscillators modeled as
\begin{align}\label{eq:eachvan1}
\begin{split}
\dot \xx_i &= \vv_i \\
\dot \vv_i &= c_i w_i (1 - \xx_i^2) \vv_i - w_i^2 \xx_i + k \uu_i, \qquad i \in \NN,
\end{split}
\end{align}
where $c_i$ and $w_i$ are the parameters for each oscillator.
Suppose that the output and the diffusive coupling input are given by
\begin{equation}\label{eq:eachvan2}
\yy_i = a \xx_i + b \vv_i \quad \text{and} \quad \uu_i = \sum_{j \in \NN_i} \alpha_{ij} (\yy_j - \yy_i)
\end{equation}
where $a > 0$ and $b>0$ are given.
For \eqref{eq:eachvan1} with \eqref{eq:eachvan2}, we claim that the synchronous and oscillatory behaviors can be achieved with a sufficiently large coupling gain $k$ if
\begin{equation}\label{eq:vdpcondi}
\frac1N \sum_{i=1}^N c_i w_i > 0 \quad \text{and} \quad \frac1N \sum_{i=1}^N w_i^2 > 0.
\end{equation}
In fact, it is well known that the stand-alone, i.e., when $\uu_i\equiv0$, Van der Pol oscillator has a stable limit cycle (and the domain of attraction is $\R^2\backslash\{0\}$) if and only if $c_iw_i>0$ and $w_i^2>0$.
Therefore, the condition \eqref{eq:vdpcondi} may be interpreted as its blended version.
Interestingly, some agent may violate $c_iw_i>0$ or $w_i^2>0$ as long as their average confirms \eqref{eq:vdpcondi}, which means that some malfunctioning oscillators may coexist in the oscillating network as long as there are a majority of good neighbors.

To justify the claim, we note that \eqref{eq:eachvan1} with \eqref{eq:eachvan2} is a heterogeneous multi-agent system, which can be rewritten as
$$\begin{bmatrix} \dot \xx_i \\ \dot \vv_i \end{bmatrix} = \begin{bmatrix} \vv_i \\ c_iw_i(1 - \xx_i^2)\vv_i - w_i^2\xx_i \end{bmatrix} - k\begin{bmatrix} 0 & 0 \\ a & b \end{bmatrix} \sum_{j=1}^{N} l_{ij} \begin{bmatrix} \xx_j \\ \vv_j \end{bmatrix}$$
where $[l_{ij}]={\mathcal L}$ is the Laplacian matrix.
This resembles the form of \eqref{eq:eachdyna}, but the coupling matrix is not symmetric.
Therefore, we employ a state transformation
$$\begin{bmatrix} \bar \xx_i \\ \bar \vv_i \end{bmatrix} = \begin{bmatrix} 1 & 0 \\ \frac{a}{b} & 1 \end{bmatrix} \begin{bmatrix} \xx_i \\ \vv_i \end{bmatrix},$$
so that the above system is converted into 
\begin{align*}
\begin{bmatrix} \dot {\bar \xx}_i \\ \dot {\bar \vv}_i \end{bmatrix} &= \begin{bmatrix} -\frac{a}{b} & 1 \\ -\begin{pmatrix} (\frac{a}b)^2 + c_iw_i\frac{a}b + w_i^2 \end{pmatrix} & \begin{pmatrix}\frac{a}b + c_iw_i \end{pmatrix}\end{bmatrix} \begin{bmatrix} \bar \xx_i \\ \bar \vv_i \end{bmatrix}\\
&\quad + c_iw_i\bar \xx_i^2 \begin{bmatrix} 0 \\ \frac{a}b \bar \xx_i - \bar \vv_i \end{bmatrix}  
- k \begin{bmatrix} 0 & 0 \\ 0 & b \end{bmatrix} \sum_{j=1}^{N} l_{ij} \begin{bmatrix} \bar \xx_j \\ \bar \vv_j \end{bmatrix} \\
&=: f_i \left(\begin{bmatrix} \bar \xx_i \\ \bar \vv_i \end{bmatrix}\right) - k B_o \sum_{j =1}^N l_{ij} \begin{bmatrix} \bar \xx_j \\ \bar \vv_j \end{bmatrix},
\end{align*}
in which $B_o$ is a positive semi-definite matrix.
This corresponds to the special case discussed in Section \ref{sec:case1}, and thus, the blended dynamics is obtained from \eqref{eq:scase1} with $Z_o = {\rm col}(1,0)$ and $W_o = {\rm col}(0,1)$ as
\begin{align}\label{eq:vanqss}
\begin{split}
	\dot{\hat{z}}_i &= \begin{bmatrix} 1 & 0 \end{bmatrix} f_i\left(\begin{bmatrix} 1 \\ 0 \end{bmatrix} \hat{z}_i + \begin{bmatrix} 0 \\ 1 \end{bmatrix} \hat{s} \right) 
	= -\frac{a}b\hat{z}_i + \hat{s}, \\
	\dot{\hat{s}} &= \frac1N  \sum_{i=1}^N \begin{bmatrix} 0 & 1\end{bmatrix} f_i\begin{pmatrix}\begin{bmatrix} \hat{z}_i \\ \hat{s} \end{bmatrix} \end{pmatrix} =: \hat f(\hat{z}_1,\dots,\hat{z}_N,\hat{s}).
\end{split}
\end{align}
Now, in order for the limiting solution to exhibit the behavior of the limit cycle, this $(N+1)$-th order blended dynamics should have a stable limit cycle, which is, however, not a trivial question in general.
However, we observe from \eqref{eq:vanqss} that, with $a/b>0$, all the $\hat z_i$'s achieve asymptotic synchronization, regardless of $\hat s$, with an exponential convergence rate.
Therefore, if the blended dynamics has a stable limit cycle, which is an invariant set, it has to be on the synchronization manifold $\mathcal{S}$ defined as
$$\mathcal{S} := \left\{ {\rm col}(\hat{z}, \dots, \hat{z}, \hat{s}) \in \mathbb{R}^{N+1} : {\rm col}(\hat{z}, \hat{s}) \in \mathbb{R}^2\right\}.$$
Projecting the blended dynamics \eqref{eq:vanqss} to the synchronization manifold $\mathcal{S}$, i.e., replacing $\hat{z}_i$ with $\hat z$ in \eqref{eq:vanqss} for all $i \in \NN$, we obtain a second-order system
\begin{align}
\dot {\hat z} &= -\frac{a}{b} \hat z + \hat s, \notag \\
\dot {\hat s} &= \hat f(\hat z, \dots, \hat z, \hat s) \label{eq:zs} \\
&= \left( \frac{a}{b} + \hat{c}\hat{w}(1 - \hat z^2)\right) \left(-\frac{a}{b} \hat z + \hat s \right) - \hat{w}^2 \hat z, \notag
\end{align}
where $\hat{w} = \sqrt{\sum_{i=1}^N w_i^2/N}$, and $\hat c$ is a constant such that $\hat{c}\hat{w} = \sum_{i=1}^N c_iw_i/N$ (this choice makes the equation look similar to the stand-alone Van der Pol oscillator).
Therefore, \eqref{eq:zs} should have a stable limit cycle if the blended dynamics has a stable limit cycle.
In order to check whether \eqref{eq:zs} actually has a stable limit cycle, define $z := \hat{z}$ and $s := -(a/b)\hat{z} + \hat{s}$, which yields
\begin{equation}\label{eq:zzss}
\dot z = s, \qquad \dot s = \hat{c}\hat{w}(1 - z^2)s - \hat{w}^2 z.
\end{equation}
This coincides with the stand-alone Van der Pol oscillator written in \eqref{eq:eachvan1}, and thus, the condition \eqref{eq:vdpcondi} is necessary and sufficient for existence of the unique stable limit cycle $\gamma$ of \eqref{eq:zzss} with the domain of attraction $\R^2 \backslash \{0\}$.
This in turn implies that \eqref{eq:vdpcondi} is a necessary condition for the blended dynamics \eqref{eq:vanqss} to have a stable limit cycle.
Further analysis, given in \citep{jingyu}, proves that the condition \eqref{eq:vdpcondi} is also sufficient for \eqref{eq:vanqss} to have a stable limit cycle.

\begin{thm1}[\cite{jingyu}]\label{thm:jg} 
	The blended dynamics \eqref{eq:vanqss} has a stable limit cycle if and only if the condition \eqref{eq:vdpcondi} holds.
	If the condition holds, the unique stable limit cycle is given as
	$$\AAA_z := \left\{{\rm col}\left(\hat z, \dots, \hat z, \frac{a}{b} \hat z + \hat s \right) \in \R^{N+1} : {\rm col}(\hat z, \hat s) \in \gamma \right\}$$
	where $\gamma \subset \mathbb{R}^2$ is the unique limit cycle of \eqref{eq:zzss}.
	Moreover, the domain of attraction of $\AAA_z$ is $\mathbb{R}^{N+1}\setminus \mathcal{M}$, where $\mathcal{M} \subset \mathbb{R}^{N+1}$ is an $(N-1)$-dimensional manifold that contains the origin and has a positive distance from $\AAA_z$; i.e., $\|m\|_{\AAA_z} \ge c > 0$, $\forall m \in {\mathcal M}$.
\end{thm1}

Theorem \ref{thm:jg} ensures that the assumption of Theorem \ref{thm:ps} is satisfied, and thus, we obtain the following.

\begin{cor1}\label{cor:van2}
	For the network of heterogeneous Van der Pol oscillators \eqref{eq:eachvan1} with \eqref{eq:eachvan2}, suppose that Assumption \ref{assum:L} and the condition \eqref{eq:vdpcondi} hold.
	Let $\DDD_x$ be defined as
	\begin{multline*}
	\DDD_x = \Big\{ {\rm col}(\xx_1, \vv_1, \dots, \xx_N, \vv_N) \in \mathbb{R}^{2N} : \\
	{\rm col}\begin{pmatrix}\xx_1, \dots, \xx_N, \frac{1}{N}\sum_{i=1}^N \left(\frac{a}{b}\xx_i + \vv_i\right)\end{pmatrix} \notin\mathcal{M} \Big\}
	\end{multline*}
	where $\mathcal{M}$ is defined in Theorem \ref{thm:jg}.
	Then, for any compact set $K \subset \DDD_x$ and for any $\eta > 0$ and $\tau >0$, there exists $k^* > 0$ such that for each $k > k^*$ and each initial condition in $K$, the solutions to \eqref{eq:eachvan1} with \eqref{eq:eachvan2} exist for all $t \ge 0$, and satisfy
	\begin{align}
    \left\|\begin{bmatrix} \xx_i(t, k) \\ \vv_i(t, k)\end{bmatrix} - \begin{bmatrix} \hat{z}_i(t) \\ - \frac{a}{b}\hat{z}_i(t) + \hat{s}(t)\end{bmatrix} - \tilde{\xi}_i(kt)\right\| &\le \eta, \,\,\, \forall t \in [0, \tau] \notag \\
	\limsup_{t \to \infty} \|{\rm col}(\xx_1, \vv_1, \dots, \xx_N, \vv_N) \|_{\AAA_x} &\le \eta, \label{eq:phasecohesive}
	\end{align}
	for all $i \in \mathcal{N}$, where $\tilde{\xi}_i$ is a function exponentially decaying to zero, and
	$$\AAA_x := \{1_N \otimes {\rm col}(z, s) : {\rm col}(z, s) \in \gamma\}.$$
\end{cor1}

Since the solution to the networked heterogeneous Van der Pol oscillator \eqref{eq:eachvan1} with \eqref{eq:eachvan2} approaches arbitrarily close to the synchronization manifold $\AAA_x$ with a sufficiently large $k$, it is seen that the solution is ``phase cohesive'' \citep[Sec.~3.1]{dorfler2014synchronization}, which means that the phase differences remain small between any two oscillators running around the limit cycle (even if the difference is not zero).
Further discussions can be found in \citep{jingyu}, where it is shown that the networked oscillator \eqref{eq:eachvan1} with \eqref{eq:eachvan2} actually has a locally asymptotically stable limit cycle whose shape becomes $\AAA_x$ as $k \to \infty$, which also implies that the solution to \eqref{eq:eachvan1} with \eqref{eq:eachvan2} converges to a periodic solution.

\begin{rem1}
	From Theorem \ref{thm:as}, we obtain 
	$$\lim_{t \to \infty} \| {\rm col}(\xx_1, \vv_1, \dots, \xx_N, \vv_N) \|_{\AAA_x} = 0$$
	for the identical case, i.e., $c_i = c_j$ and $w_i = w_j$ for any $i \neq j$.
	This is because the identical case guarantees the condition (1) of Theorem \ref{thm:as}, and the limit cycle $\AAA_z$ is proved to be locally exponentially stable in \citep{jingyu}.
\end{rem1}

\subsection{Estimation of the Number of Agents}\label{sec:application0}

When constructing a distributed network, sometimes there is a need for each agent to know global information such as the number of agents in the network without resorting to a centralized unit.
See \citep{Kos05,Baq12} for more on this problem.
In such circumstances, the proposed tool can be employed to design a distributed network that estimates the number of participating agents, under the assumption that there is one agent (whose ID is 1, for example) who always takes part in the network.
Suppose that agent $1$ integrates the following scalar dynamics:
\begin{equation}\label{eq:N1}
\dot{n}_1(t) = -n_1(t) + 1 + k \sum_{j \in \NN_1} (n_j(t) - n_1(t))
\end{equation}
while all others integrate
\begin{equation}\label{eq:N2}
\dot{n}_i(t) = 1 + k\sum_{j \in \NN_i} (n_j(t) - n_i(t)), \;\;\; i = 2, \dots, N
\end{equation}
where $N$ is unknown to the agents.
Then, the multi-agent system is heterogeneous and corresponds to the special case of Section \ref{sec:case3}.
Therefore, the blended dynamics is simply obtained as
\begin{equation}\label{eq:Ns}
\dot{\hat s}(t) = -\frac{1}{N}\hat{s}(t) + 1.
\end{equation}
This implies that all the limiting solutions $\xi_i(t)=\hat{s}(t)$ converge to $N$ as time goes to infinity.
Then, it follows from Theorem \ref{thm:contraction} that each state $n_i(t)$ approaches arbitrarily close to $N$ with a sufficiently large $k$.
Hence, by increasing $k$ such that the estimation error is less than $0.5$, and by rounding $n_i(t)$ to the nearest integer, each agent gets to know the number $N$ as time goes on.
As a matter of fact, if there is a known upper bound $N_{\max}$ of $N$, one can obtain an explicit result as follows.

\begin{cor1}[\cite{donggil}]
Suppose that Assumption \ref{assum:L} hold, the agent $1$ always participates in the network, and there is a known upper bound $N_{\max}$ of $N$.
Then, for each $k > k^* = N_{\max}^3$ and $n_i(0) \in [0, N_{\max}]$, $i\in\NN$, the solution to \eqref{eq:N1} and \eqref{eq:N2} exists for all $t \ge 0$, and satisfies
$$\left\lceil n_i(t) \right\rfloor = N, \quad \forall i\in \mathcal{N}, \quad \forall t \ge T$$
where $T = 4N_{\max}\ln(2N_{\max}^{1.5}k/(k-N_{\max}^3))$, and $\lceil \cdot \rfloor$ is the rounding operator.
\end{cor1}

\begin{rem1}
	A slight variation of the idea can also yield an algorithm to identify the agents attending the network.
	Let the number $1$ in both \eqref{eq:N1} and \eqref{eq:N2} be replaced by $2^{i-1}$, where $i$ is the unique ID of the agent in $\{1, 2, \dots, N_{\max}\}$.
	Then the blended dynamics \eqref{eq:Ns} becomes $\dot {\hat s} = -(1/N) \hat s + \sum_{j \in \NN_a} 2^{j-1}/N$, where $\NN_a$ is the index set of the attending agents, and $N$ is the cardinality of $\NN_a$.
	Since the limiting solution $\xi_i(t) \to \sum_{j \in \NN_a} 2^{j-1}$, each agent can figure out the integer value $\sum_{j \in \NN_a} 2^{j-1}$, which contains the binary information of the attending agents.
\end{rem1}

The drawbacks of the proposed algorithm are those two assumptions that the agent $1$ should always join the network, and the maximum number $N_{\max}$ should be known.
On the other hand, one benefit of the proposed method is that the initial conditions do not affect the final value of $n_i(t)$ because they are forgotten as time tends to infinity due to the stability of the blended dynamics.
This is in sharp contrast to other works such as \citep{shames2012distributed}, where the information is encoded in the initial conditions, which is then sensitive to disturbance or noise.
In this work, we embed the information in the vector fields (such as $-n_1(t)+1$ in \eqref{eq:N1} or $1$ in \eqref{eq:N2}), and rely on the stability to make each agent converge to a computed outcome.
This is a method that is robust against disturbance or noise.
In addition, since it does not rely on the initial conditions (we call it initialization-free), the proposed algorithm is suitable for plug-and-play operation; e.g., some agents may join or leave the network during the operation.
Further discussions are found in \citep{donggil}, which illustrates how the plug-and-play operation is guaranteed for the proposed algorithm of \eqref{eq:N1} and \eqref{eq:N2}.

\subsection{Distributed Optimization}\label{sec:application3}

Consider an optimization problem 
\begin{align}\label{eq:opt_prob}
\text{minimize } &\sum_{i=1}^{N} J_i(x_i) \nonumber\\
\text{subject to } &\sum_{i=1}^N x_i = \sum_{i=1}^N d_i, \;\; \underline x_i \le x_i \le \overline x_i, \;\; i \in \NN
\end{align}
where $x_i \in \R$ is the decision variable, $J_i$ is a strictly convex $C^2$ function, and $\underline x_i$, $\overline x_i$, and $d_i$ are given constants.
A practical example is the economic dispatch problem of electric power, in which $d_i$ represents the demand of node $i$, $x_i$ is the power generated at node $i$ with its minimum $\underline x_i$ and maximum $\overline x_i$, and $J_i$ is the generation cost.

A centralized solution is easily obtained using Lagrangian and Lagrange dual functions.
Indeed, it can be shown that the optimal value is obtained by $x_i^* = \theta_i(\lambda^*)$ where
$$\theta_i(\lambda) := \left(\frac{dJ_i}{dx_i}\right)^{-1}\left(\text{sat} \left(\lambda, \frac{dJ_i}{dx_i}(\underline{x}_i), \frac{dJ_i}{dx_i}(\overline{x}_i)\right) \right),$$
where $(dJ_i/dx_i)^{-1}(\cdot)$ is the inverse function of $(dJ_i/dx_i)(\cdot)$, $\text{sat}(s,a,b)$ is $s$ if $a \le s \le b$, $b$ if $b < s$, $a$ if $s<a$.
The optimal $\lambda^*$ maximizes the dual concave function $g(\lambda) = \sum_{i=1}^N J_i(\theta_i(\lambda)) + \lambda (d_i - \theta_i(\lambda))$, which can be asymptotically obtained by the gradient algorithm:
\begin{equation}\label{eq:cen_edp}
\dot{\lambda}(t) = \frac{dg}{d\lambda}(\lambda(t)) = \sum_{i=1}^N (d_i - \theta_i(\lambda(t))) .
\end{equation}\vspace{-5mm}

On the other hand, a distributed algorithm to solve the optimization problem approximately is to integrate
\begin{equation}\label{eq:do}
\dot{\lambda}_i(t) = d_i - \theta_i(\lambda_i(t)) + k \sum_{j \in \NN_i} (\lambda_j(t) - \lambda_i(t))
\end{equation}
at each node $i \in \NN$.
Note that the function $\theta_i$ can be computed within the node $i$ from local information such as $J_i$, $\underline x_i$, and $\overline x_i$, and thus, the proposed solver \eqref{eq:do} does not exchange the private information of each node with other nodes (except the dual variable $\lambda_i$).
The idea of constructing \eqref{eq:do} comes from the fact that the blended dynamics of the heterogeneous multi-agent system \eqref{eq:do} is given by
\begin{equation}\label{eq:bdo}
\dot{\hat s}(t) = \frac{1}{N} \sum_{i=1}^N (d_i - \theta_i(\hat s(t))) = \frac1N \frac{dg}{d\lambda}(\hat s(t)) 
\end{equation}
which follows from Section \ref{sec:case3}.
Obviously, \eqref{eq:bdo} is the same as the centralized solver \eqref{eq:cen_edp} except the scaling of $1/N$.
The effect of this scaling can be compensated considering that the time $t$ in \eqref{eq:bdo} is not a physical time but computer time, which can be accelerated in a faster computer.
Finally, by Theorem \ref{thm:contraction}, the state $\lambda_i(t)$ of each node approaches arbitrarily close to $\lambda^*$ with a sufficiently large $k$, and so, the following is obtained.

\begin{cor1}
Suppose that Assumption \ref{assum:L} holds and the optimization problem \eqref{eq:opt_prob} is feasible.
Then, for any $\eta > 0$, there exist $k^* > 0$ and $T > 0$ such that, for each $k > k^*$ and $\lambda_i(0) \in [dJ_i/dx_i(\underline{x}_i), dJ_i/dx_i(\overline{x}_i)]$, $i \in \NN$, the solution to \eqref{eq:bdo} exists for all $t \ge 0$, and satisfies
$$\|\theta_i(\lambda_i(t)) - x_i^*\| \le \eta, \quad \forall i \in \mathcal{N}, \quad \forall t \ge T$$
where $x_i^*$ is the optimal solution.
\end{cor1}

Readers are referred to \citep{yun2018initialization}, which also describes the behavior of the proposed algorithm when the problem is infeasible so that each agent can figure out that infeasibility occurs.
It is again emphasized that the initial conditions are forgotten and therefore unimportant, which makes the algorithm suitable for plug-and-play operation.

\subsection{A Remark on the Coupling Gain $k$}

In general, computation of the threshold $k^*$ for the coupling gain $k$ requires the information about the agent dynamics, the graph Laplacian, and so on, which may become an obstacle for the decentralized design of multi-agent systems, even though the operation of the multi-agent system is still distributed.
This is not a problem if a designer computes $k^*$ a priori and broadcasts $k$ to every agent before the operation.
If not, some other ideas such as a worst-case selection of $k^*$ or adaptive on-line tuning of $k$ may need to be employed. 
For more details, one can refer to the papers cited in the previous subsections.

\section{Conclusion}\label{sec:conclusion}

We have illustrated that the behavior of a heterogeneous multi-agent system under rank-deficient strong coupling can be approximated by the limiting solution that is obtained from the blended dynamics.
This may serve as a useful toolkit for the analysis of multi-agent systems whose heterogeneity comes from the uncertainty in nature, or for the synthesis of networked control systems whose heterogeneity is intentionally posed in order for a different agent to perform a different task.
Various properties of the proposed approach are discussed in the introduction, and several applications are included to emphasize these properties.

For the future works, a possible direction will be to consider nonlinear coupling as in \citep{Park10,Sepulchre11}, communication disturbance as in \citep{And18}, and communication delay as in \citep{Hatanaka16}.
Consideration of different coupling gains, rather than the common $k$, should also be of interest.
Finally, we recall that the proposed analysis is based on the idea of treating the coupling gain $k$ as the singular perturbation parameter that is sufficiently large.
However, for synthesis, there can be a limit for increasing $k$ because of discretization, physical saturation, and so on.
The study in this paper has a limitation for these cases.

\appendix

\section{Proofs}\label{sec:proof}

In order to prove all the claims for the two cases, \eqref{eq:eachdyna} and \eqref{eq:eachdynb}, in a compact way, let us define a new variable $\ZZ := {\rm col}(z_1, \dots, z_N, z_o)$.
Then, both systems \eqref{eq:tsysa} and \eqref{eq:tsysb} can be uniformly written as 
\begin{align}\label{eq:tran}
\begin{split}
\dot \ZZ &= h(t, \ZZ, w) \\
\epsilon \dot w &= \epsilon g(t, \ZZ, w) - Qw 
\end{split}
\end{align}
where $\ZZ \in \R^m$, $w \in \R^{nN-m}$, $\epsilon = 1/k$, both $h$ and $g$ are defined from \eqref{eq:tsysa} or \eqref{eq:tsysb}, which are piecewise continuous in $t$, continuously differentiable with respect to $\ZZ$ and $w$, locally Lipschitz with respect to $\ZZ$ and $w$ uniformly in $t$, and both $h(t,0,0)$ and $g(t, 0, 0)$ are uniformly bounded in $t$.
In fact, ${\rm col}(\ZZ,w) = P_{a,b} {\rm col}(x_1,\dots,x_N)$
where $P_{a,b}$ denotes the matrix $P_a$ or $P_b$ in \eqref{eq:Pa} or \eqref{eq:Pb} depending on the case.
Therefore, the initial conditions $\ZZ(0)$ and $w(0)$ are determined from $x(0)$.
If a compact set $K \subset \R^{nN}$ is given for the initial condition $x(0)$, then let $K_{\ZZ} \subset \R^m$ and $K_w \subset \R^{nN-m}$ be compact sets such that, for all $x(0) \in K$, the corresponding $\ZZ(0) \in K_{\ZZ}$ and $w(0) \in K_w$.
Now, the quasi-steady-state subsystem of \eqref{eq:tran} can be denoted by
\begin{equation}\label{eq:red}
\dot {\hat \ZZ} = h(t, \hat{\ZZ}, 0)
\end{equation}
which corresponds to the blended dynamics \eqref{eq:qss}, and it is noted that ${\rm col}(\hat \ZZ,0) = P_{a,b} \xi$ where $\xi := {\rm col}(\xi_1,\dots,\xi_N)$.
Its initial condition satisfies $\hat \ZZ(0) =~\!\ZZ(0)$.

The following lemma, whose statement is customized to our case, will be used frequently in the proofs.

\begin{lem1}[\cite{khalil2002nonlinear}, Theorem 11.1]\label{lem:tykh}
Assume that there is a compact set $K_\ZZ$ such that, for any $\hat \ZZ(0) \in K_\ZZ \subset \R^m$, the solution $\hat \ZZ(t)$ to \eqref{eq:red} exists for all $t \ge 0$.
Then, for any compact set $K_w \subset \R^{nN-m}$, and for any $\eta'>0$ and $\tau'>0$, there exists $\epsilon^* > 0$ such that, for each $0 < \epsilon < \epsilon^*$ and $w(0) \in K_w$, the solution to \eqref{eq:tran} with $\ZZ(0)=\hat \ZZ(0)$ exists for $t \in [0,\tau']$ and satisfies
\begin{align}\label{eq:applemma}
\begin{split}
\| \ZZ(t) - \hat\ZZ(t) \| &\le \eta', \quad \forall t \in [0,\tau'], \\
\| w(t) - \exp(-Qt/\epsilon)w(0) \| &\le \eta', \quad \forall t \in [0,\tau'].
\end{split}
\end{align}
\end{lem1}

\subsection{Proof of Proposition \ref{prop:extreme}}

The claim of Proposition \ref{prop:extreme} is proved by showing that for arbitrarily small $\eta>0$, there exists $\epsilon^*>0$ such that, for all $k > 1/\epsilon^*$, we have $\|\ZZ(t,k) - \hat \ZZ(t)\| \le \eta/2$ and $\|w(t,k)\| \le \eta$ at given $t>0$, where $(\ZZ(t,k),w(t,k))$ is the solution to \eqref{eq:tran} with $\epsilon=1/k$ and given initial condition $(\ZZ(0),w(0))$.
For this, set $\eta' = \eta/2$, $\tau' = t$, $K_\ZZ = \{\ZZ(0)\}$, $K_w = \{w(0)\}$, and apply Lemma \ref{lem:tykh} to obtain $\epsilon^*$ for \eqref{eq:applemma}. Note that the assumption of Proposition \ref{prop:extreme} implies the assumption of Lemma \ref{lem:tykh}.
Now, without loss of generality, suppose that $\epsilon^*$ is small enough so that $\|\exp(-Q \tau'/\epsilon)w(0)\| \le \eta/2$ for all $\epsilon \in(0, \epsilon^*)$.
Then, \eqref{eq:applemma} at $t=\tau'$ justifies the claim.

\subsection{Proof of Theorem \ref{thm:contraction}}\label{app:cont}

From the assumption of Theorem \ref{thm:contraction}, there exist a positive definite matrix $H_c$ and a positive constant $\lambda_c$ such that 
\begin{align*}
&(\ZZ_1 - \ZZ_2)^TH_c(h(t, \ZZ_1, 0) - h(t, \ZZ_2, 0)) \\
&= (\ZZ_1 - \ZZ_2)^TH_c\frac{\partial h}{\partial \ZZ}\left(t, r\ZZ_1 + (1-r)\ZZ_2, 0\right)(\ZZ_1 - \ZZ_2) \\
&=\frac{1}{2}(\ZZ_1 - \ZZ_2)^T\left[H_c\left(\frac{\partial h}{\partial \ZZ}\right) + \left(\frac{\partial h}{\partial \ZZ}\right)^TH_c\right](\ZZ_1 - \ZZ_2) \\
&\le -\frac{\lambda_c}{2}(\ZZ_1 - \ZZ_2)^TH_c(\ZZ_1 - \ZZ_2)
\end{align*}
with some $r \in (0, 1)$ for each $\ZZ_1$ and $\ZZ_2 \in \mathbb{R}^m$.
With this in mind, let us define $\VV_c(\hat{\ZZ}, \ZZ) := (\hat \ZZ - \ZZ)^T H_c (\hat \ZZ - \ZZ)$ and $\VV_c^0(\hat \ZZ) := \VV_c(\hat \ZZ,0)$.
Then, we claim that the solution $\hat \ZZ(t)$ of \eqref{eq:red} with arbitrary $\hat \ZZ(0) \in K_{\ZZ}$, is bounded so that there is $\mu>0$ such that $\hat \ZZ(t) \in \Omega_c^0(\mu) := \{ \hat \ZZ : \VV_c^0(\hat \ZZ) \le \mu \}$ for all $t \ge 0$.
Indeed, since
\begin{align*}
\dot{\VV}_c^0 &= 2\hat{\ZZ}^TH_c(h(t, \hat{\ZZ}, 0) - h(t, 0, 0)) + 2\hat{\ZZ}^TH_ch(t, 0, 0) \\
&\le -\lambda_c\VV_c^0 + 2\sqrt{\bar{h}}\sqrt{\VV_c^0}
\end{align*}
where $\bar h := \sup_{t \ge 0}\VV_c^0(h(t, 0, 0))$, we have $\dot \VV_c^0 \le 0$ whenever $\VV_c^0(\ZZ) \ge 4\bar h/\lambda_c^2$.
Therefore, by taking $\mu$ such that $\mu \ge 4 \bar h/\lambda_c^2$ and $\Omega_c^0(\mu) \supset K_{\ZZ}$, the claim is justified.

Now, let $\bar \VV_c(\hat \ZZ,\ZZ,w) := \VV_c(\hat \ZZ,\ZZ) + w^T H w$ where $H$ is the positive definite matrix satisfying $Q^T H + H Q = I$.
Define $\Omega_c(\bar \eta) := \{ (\hat \ZZ,\ZZ,w) : \bar \VV_c(\hat \ZZ,\ZZ,w) \le \bar \eta\}$.
For a given $\eta$, set 
$$\bar{\eta} = \frac{\eta^2}{2\|P_{a,b}^{-1}\|^2}\min\{\lambda_{\text{min}}(H_c), \lambda_{\text{min}}(H)\}.$$
Then, $(\hat \ZZ, \ZZ, w) \in \Omega_c(\bar\eta)$ implies $\|P_{a, b}^{-1}\|^2\|\hat \ZZ - \ZZ\|^2 \le \eta^2/2$ and $\|P_{a,b}^{-1}\|^2\|w\|^2 \le \eta^2/2$.
This, in turn, implies $\|x_i-\xi_i\| \le \eta$ because $\|x_i - \xi_i\|^2 \le \|x-\xi\|^2 = \|P_{a,b}^{-1} ({\rm col}(\ZZ, w) - {\rm col}(\hat{\ZZ}, 0))\|^2 = \|P_{a,b}^{-1}\|^2 (\|\ZZ - \hat{\ZZ}\|^2 + \|w\|^2) \le \eta^2$.
Therefore, the remaining proof shows that there exists $\epsilon^*>0$ such that, for each $\epsilon \in (0,\epsilon^*)$, (i) the set $\Omega := \Omega_c(\bar \eta) \cap (\Omega_c^0(\mu) \times \mathbb{R}^{nN})$, which is compact, is positively invariant and (ii) the solution $(\hat \ZZ(t),\ZZ(t),w(t))$ of both \eqref{eq:tran} and \eqref{eq:red} enters the set $\Omega$ in a finite time.

(i): The time derivative of $\bar{\VV}_c$ for both \eqref{eq:tran} and \eqref{eq:red} on the compact set $\Omega$ yields that
\begin{align*}
&\dot{\bar{\VV}}_c = 2(\ZZ - \hat{\ZZ})^TH_c(h(t, \ZZ, 0) - h(t, \hat{\ZZ}, 0)) - \frac{1}{\epsilon}\|w\|^2 \\
&+2(\ZZ - \hat{\ZZ})^TH_c(\partial h/\partial w)(t, \ZZ, r w)w +2w^THg(t, \ZZ, w) \\
&\le -\lambda_c\VV_c - \frac{1}{\epsilon}\|w\|^2 + 2\bar{\theta}\sqrt{\|H_c\|}\sqrt{\VV_c}\|w\| + 2\bar{\theta}\|H\|\|w\| \\
&\le -\frac{\lambda_c}{2}\bar{\VV}_c + 2\epsilon\bar{\theta}^2\|H\|^2 - \begin{bmatrix}\frac{1}{2\epsilon} - \frac{2\bar{\theta}^2\|H_c\|}{\lambda_c} - \frac{\lambda_c\|H\|}{2}\end{bmatrix}\|w\|^2
\end{align*}
where $\bar{\theta}$ is chosen such that $\bar \theta$ is an upper bound of $\|(\partial h/\partial w)(t,\ZZ,w)\|$ and $\|g(t,\ZZ,w)\|$ for all $t \ge 0$ on the set $\Omega$.
Take $\bar{\epsilon} = \lambda_c \min\{ \bar\eta/(4\bar{\theta}^2\|H\|^2), 1/(4\bar \theta^2 \|H_c\| + \lambda_c^2 \|H\|) \}$.
Then, for each $\epsilon \in (0,\bar\epsilon)$, $\dot {\bar \VV}_c \le 0$ if $\bar \VV_c \ge \bar \eta$.
Hence, the set $\Omega$ is positively invariant for such $\epsilon$. 

(ii): Set $\eta' = \sqrt{\bar{\eta}/(\|H_c\| + 4\|H\|)}$, and find $\tau'>0$ such that, for any $w(0) \in K_w$, we have $\|\exp(-Qt/\epsilon)w(0)\| \le \eta'$ for all $t \ge \tau'$ and all $\epsilon \in (0, \bar{\epsilon})$.
Then, with given $K_{\ZZ}$ and $K_w$, Lemma \ref{lem:tykh} yields $\epsilon^*$ (such that $\epsilon^* \le \bar \epsilon$ without loss of generality).
Then, for each $\epsilon \in (0,\epsilon^*)$, the solutions satisfy $\|\ZZ(\tau') - \hat \ZZ(\tau')\| \le \eta'$ and $\|w(\tau')\| \le 2\eta'$, which guarantees that $\bar \VV_c(\hat \ZZ(\tau'),\ZZ(\tau'),w(\tau')) \le \bar \eta$.

\subsection{Proof of Theorem \ref{thm:ps}}\label{app:ps}

The proof for the inequality \eqref{eq:setconv} of Theorem \ref{thm:ps} is a direct consequence of Lemma \ref{lem:tykh} because it is just a different expression of \eqref{eq:applemma} via the linear coordinate change $P_{a,b}$.

Now, the assumption of Theorem \ref{thm:ps} implies that the set $\AAA_z$ is uniformly asymptotically stable for \eqref{eq:red} and that there exists $r_0 > 0$ such that $\AAA_z \subset \AAA_z^{r_0} := \{ \ZZ : \|\ZZ\|_{\AAA_z} \le r_0 \} \subset \DDD_z$.
Then, there exist a continuously differentiable function $\VV_\ZZ : \AAA_z^{r_0} \times [0, \infty) \to \mathbb{R}_{\ge 0}$, class $\mathcal{K}$ functions $\alpha_1$, $\alpha_2$, and $\alpha_3$, and a constant $\theta$ such that
\begin{gather}
\alpha_1(\|\ZZ\|_{\AAA_z}) \le \VV_\ZZ(\ZZ, t) \le \alpha_2(\|\ZZ\|_{\AAA_z}) \notag \\
\frac{\partial \VV_\ZZ}{\partial t}(\ZZ, t) + \frac{\partial \VV_\ZZ}{\partial \ZZ}(\ZZ, t) h(t, \ZZ, 0) \le -\alpha_3(\VV_\ZZ(\ZZ, t)) \notag \\
\left\| \frac{\partial \VV_\ZZ}{\partial \ZZ}(\ZZ, t)\right\| \le \theta \label{eq:eqconv}
\end{gather}
for all $\ZZ \in \AAA_z^{r_0}$ and $t \ge 0$ (see Appendix \ref{app:conv} for the construction of $\VV_\ZZ$).
Let $\bar{\VV}_\ZZ(\ZZ,w,t) := \VV_\ZZ(\ZZ, t) + w^T H w$ where $H$ is the positive definite solution to $Q^TH+HQ=I$.
For a given $\eta$, let
$$\bar{\eta} := \min\begin{Bmatrix}\alpha_1 \left(\frac{\eta}{\sqrt{2}\|P_{a,b}^{-1}\|}\right), \,\, \frac{\eta^2\lambda_{\text{min}}(H)}{2\|P_{a,b}^{-1}\|^2}, \,\, \alpha_1(r_0) \end{Bmatrix}$$
and let $\Omega(t) := \{ (\ZZ,w) : \bar \VV_\ZZ(\ZZ,w,t) \le \bar \eta \}$.
Then, with $\eta' := \min\{ \alpha_2^{-1}(\bar{\eta}/2)/2, \sqrt{\bar{\eta}/(2\|H\|)}/2 \}$ and $\eta'' := \eta/(\sqrt{2}\|P_{a,b}^{-1}\|)$ for given $\eta$, it can be shown that $\Omega_{\rm inner} := \{ (\ZZ,w) : \|\ZZ\|_{\AAA_z} \le 2\eta', \|w\| \le 2\eta' \} \subset \Omega(t) \subset \Omega_{\rm outer} := \{ (\ZZ,w) : \|\ZZ\|_{\AAA_z} \le \min\{ \eta'', r_0\}, \|w\| \le \eta'' \}$ for all $t \ge 0$ ($2\eta' < \eta''$).
On the other hand, $(\ZZ,w) \in \Omega_{\rm outer}$ implies $\|x\|_{\AAA_x} \le \eta$ because $\|x\|_{\AAA_x}^2 = \inf_{\hat{\ZZ} \in \AAA_z} \| P_{a,b}^{-1} {\rm col}(\ZZ, w) - P_{a,b}^{-1} {\rm col}(\hat{\ZZ}, 0) \|^2 = \|P_{a,b}^{-1}\|^2 (\|\ZZ\|_{\AAA_z}^2 + \|w\|^2)$.
Therefore, the following proof shows that there exists $\epsilon^*>0$ such that, for each $\epsilon \in (0,\epsilon^*)$, (i) $\dot {\bar \VV}_\ZZ \le 0$ for all $t \ge 0$ on the set $\Omega_{\rm outer} \setminus \Omega_{\rm inner} \subset \AAA_z^{r_0} \times \{ w: \|w\| \le \eta''\}$, and (ii) there is a time $\tau' \ge 0$ at which $(\ZZ(\tau'),w(\tau')) \in \Omega_{\rm inner}$. 

(i): The time derivative of $\bar \VV_\ZZ$ along \eqref{eq:tran} on $\Omega_{\rm outer} \setminus \Omega_{\rm inner}$ leads to
\begin{align*}
\dot{\bar{\VV}}_\ZZ &= \frac{\partial \VV_\ZZ}{\partial t}(\ZZ, t) + \frac{\partial \VV_\ZZ}{\partial \ZZ}(\ZZ, t)h(t, \ZZ, 0) - \frac{1}{\epsilon}\|w\|^2 \\
& + \frac{\partial \VV_\ZZ}{\partial \ZZ}(\ZZ, t)(\partial h/\partial w)(t, \ZZ, rw)w + 2w^THg(t, \ZZ, w) \\
&\le -\alpha_3(\VV_\ZZ(\ZZ, t)) - \frac{1}{2\epsilon}\|w\|^2 + 2\epsilon \bar \theta^2 \\
&\le -\min\left\{\alpha_3(\alpha_1(2\eta')),\frac{2}{\epsilon}(\eta')^2\right\} + 2\epsilon \bar\theta^2
\end{align*}
in which, $\bar\theta$ is chosen such that $\bar\theta$ is an upper bound of $\theta \|(\partial h/\partial w)(t,\ZZ,w)\|$ and $2 \|H\| \|g(t,\ZZ,w)\|$ for all $t \ge 0$ on $\Omega_{\rm outer} \setminus \Omega_{\rm inner}$.
Take $\bar{\epsilon} = \min\{\alpha_3(\alpha_1(2\eta'))/(2\bar\theta^2), \eta'/\bar\theta\}$.
Then, for each $\epsilon \in (0, \bar{\epsilon})$, $\dot {\bar \VV}_\ZZ \le 0$ by the above inequality.

(ii): Since the set $\AAA_z$ is uniformly asymptotically stable for \eqref{eq:red}, there exists $\tau'$ such that $\|\hat{\ZZ}(t)\|_{\AAA_z} \le \eta'$ for all $t \ge \tau'$ and that $\|\exp(-Q t/\bar{\epsilon}) w(0)\| \le \eta'$ for all $t \ge \tau'$ and all $w(0) \in K_w$.
By applying Lemma \ref{lem:tykh} with $\eta'$, $\tau'$, and $K_w$, we obtain $\epsilon^*>0$ (without loss of generality, $\epsilon^* \le \bar \epsilon$), such that $\|\ZZ(\tau')\|_{\AAA_z} \le \|\hat\ZZ(\tau')\|_{\AAA_z} + \|\ZZ(\tau') - \hat{\ZZ}(\tau')\| \le 2 \eta'$ and $\|w(\tau')\| \le \eta' + \|\exp(-Q \tau'/\epsilon) w(0)\| \le 2\eta'$ for all $\epsilon \in (0,\epsilon^*)$, i.e., $(\ZZ(\tau'),w(\tau')) \in \Omega_{\rm inner}$.

\subsection{Proof of Theorem \ref{thm:as}}

Now, in order to prove Theorem \ref{thm:as}, we note that, by the condition (2) of Theorem \ref{thm:as}, there is a Lyapunov function $\UU$ such that $\alpha_1 \|\ZZ\|_{\AAA_z}^2 \le \UU(\ZZ,t) \le \alpha_2 \|\ZZ\|_{\AAA_z}^2$, $\|(\partial \UU/\partial \ZZ)(\ZZ,t)\| \le \alpha_3 \|\ZZ\|_{\AAA_z}$, and $(\partial \UU/\partial t)(\ZZ,t) + (\partial \UU/\partial \ZZ)(\ZZ,t) h(t,\ZZ,0) \le -\alpha_4 \|\ZZ\|_{\AAA_z}^2$, with some $\alpha_i>0$, that is valid on a neighborhood $\Omega^0 := \{\ZZ : \alpha_1\|\ZZ\|_{\AAA_z}^2 \le \bar{\eta}\}$ with some $\bar{\eta}>0$.
Let $\bar\UU(\ZZ, w, t) := \UU(\ZZ,t) + w^THw$ where $H$ is the positive definite matrix such that $Q^TH + HQ = I$.
The condition (1) of Theorem \ref{thm:as} implies that $g(t, \ZZ, 0) = 0$ for all $\ZZ \in \AAA_z$ and $t \ge 0$, and thus, there is a constant $\bar g$ such that $\|Hg(t, \ZZ, 0)\| \le \bar g \|\ZZ\|_{\AAA_z}$ on the set $\Omega^0$.
Moreover, we can find constants $\theta_{\bar{h}}$ and $\theta_{\bar{g}}$ such that $\|\alpha_3(\partial h/\partial w)\| \le \theta_{\bar{h}}$ and $\|2H(\partial g/\partial w)\| \le \theta_{\bar{g}}$ on the compact set $\Omega := \Omega^0 \times \{ w : w^THw \le \bar\eta\}$ for all $t \ge 0$.
Then, the time derivative of $\bar{\UU}$ along \eqref{eq:tran} on $\Omega$ satisfies
\begin{align*}
\dot{\bar{\UU}} &= \frac{\partial \UU}{\partial t}(\ZZ, t) + \frac{\partial \UU}{\partial \ZZ}(\ZZ, t)h(t, \ZZ, 0) -\frac{1}{\epsilon}\|w\|^2 \\
&\quad +\frac{\partial \UU}{\partial \ZZ}(\ZZ, t)(\partial h/\partial w)(t, \ZZ, r_1w)w + 2w^THg(t, \ZZ, 0) \\
&\quad + 2w^TH(\partial g/\partial w)(t, \ZZ, r_2w)w \\
&\le -\alpha_4\|\ZZ\|_{\AAA_z}^2 + (\theta_{\bar{h}} + 2\bar{g})\|\ZZ\|_{\AAA_z}\|w\| - \begin{bmatrix}\frac{1}{\epsilon} - \theta_{\bar{g}}\end{bmatrix}\|w\|^2.
\end{align*}
Hence, we can find $\bar \epsilon > 0$ and $\lambda_u > 0$ such that, for all $\epsilon \in (0, \bar \epsilon)$ and $t \ge 0$, we have $\dot{\bar \UU} \le -\lambda_u\bar \UU$ whenever $\bar \UU \le \bar \eta$ (for each $t \ge 0$, $\{(\ZZ, w) : \bar \UU(\ZZ, w, t) \le \bar \eta\} \subset \Omega$).

Let $\eta' := \sqrt{\bar\eta/(4(\alpha_2 + \|H\|))}$, then we obtain as in (ii) of Appendix \ref{app:ps}, $\tau'$ and $\epsilon^* > 0$ ($\epsilon^* \le \bar{\epsilon}$), such that $\bar\UU(Z, w, \tau') \le \alpha_2\|\ZZ\|_{\AAA_z}^2 + w^THw \le \bar \eta$ for all $\epsilon \in (0, \epsilon^*)$.
This implies $\lim_{t \to \infty} \|x(t, k)\|_{\AAA_x} = 0$, which completes the proof with $k^{*}:=1/\epsilon^{*}$.

\section{Converse Lyapunov Theorem for a Set}\label{app:conv}

We provide with this version of converse Lyapunov theorem because we were not able to find it for the class of time-varying systems having a compact attractor that is locally asymptotically stable, and we also need \eqref{eq:eqconv}.

Note that there exists $r_0 > 0$ such that $\{\ZZ: \|\ZZ\|_{\AAA_z}\le r_0\} \subset \DDD_z$.
Then, by the uniform stability, there is a class $\mathcal{K}$ function $\delta$ such that for any $\eta \in (0, r_0]$, the solution of \eqref{eq:red} with $\hat{\ZZ}(t_0) = \ZZ_0$, denoted as $\hat{\ZZ}(t, t_0, \ZZ_0)$, satisfies
$$\|\hat{\ZZ}(t, t_0, \ZZ_0)\|_{\AAA_z} \le \eta, \; \forall t_0 \ge 0, \, \forall t \ge t_0, \, \forall \|\ZZ_0\|_{\AAA_z} \le \delta(\eta).$$
Moreover, given $\eta \in (0, r_0]$ and $r \in (0, \delta(r_0)]$, there exists $T(\eta, r) \ge 0$ such that $\|\hat{\ZZ}(t, t_0, \ZZ_0)\|_{\AAA_z} \le \eta$ for all $t_0 \ge 0$, $t \ge T(\eta, r) + t_0$, and $\|\ZZ_0\|_{\AAA_z} \le r$.
Then, by following the proof of \cite[Lemma 4.5]{khalil2002nonlinear}, we obtain a class $\mathcal{KL}$ function $\beta: [0,\delta(r_0)] \times [0,\infty) \to \R_{\ge 0}$ satisfying
$$\|\hat{\ZZ}(t, t_0, \ZZ_0)\|_{\AAA_z} \le \beta(\|\ZZ_0\|_{\AAA_z}, t - t_0)$$
for all $t_0 \ge 0$, $t \ge t_0$, and $\|\ZZ_0\|_{\AAA_z} \le \delta(r_0)$.
Now, by \cite[Lemma 15]{teel2000smooth}, we obtain a function $\omega: \{\ZZ : \|\ZZ\|_{\AAA_z} < r_0 + 1\} =: O \to \mathbb{R}$ that is continuous on $O$ and smooth on $O\setminus \AAA_z$, which satisfies, for all $\hat{\ZZ} \in O$, 
$$0.5\|\hat{\ZZ}\|_{\AAA_z} \le \omega(\hat{\ZZ}) \le 1.5\|\hat{\ZZ}\|_{\AAA_z}.$$
Then, by \cite[Lemma 17]{teel2000smooth}, we get $\rho \in \mathcal{K}_{\infty}$ such that $\hat{\omega}(\cdot) := \rho(\omega(\cdot))$ is smooth on $O$.

Because $h(t, \ZZ, 0)$ is $C^1$ and locally Lipschitz with respect to $\ZZ$ uniformly in $t$, we can find a constant $L$ such that $\left\| (\partial h/ \partial \ZZ)(t, \ZZ, 0)\right\| \le L$ for all $\|\ZZ\|_{\AAA_z}\le r_0$ and $t \ge 0$. Let $\delta_0 := \delta(r_0)/2$, then we obtain, as in \cite[Appendix C.7]{khalil2002nonlinear}, a continuously differentiable function $\VV_\ZZ : \Omega_\ZZ \times [0, \infty) \to \mathbb{R}_{\ge 0}$ as
$\VV_\ZZ(\ZZ, t) := \int_{t}^{\infty} G(\hat{\omega}(\hat{\ZZ}(\tau, t, \ZZ))) d\tau$, where $G$ is a class $\mathcal{K}$ function obtained from \cite[Lemma C.1]{khalil2002nonlinear}, by letting $g(s) = \rho(1.5\beta(\delta_0, s))$ and $h(s) = e^{Ls}$.
The properties of the Lyapunov function listed around \eqref{eq:eqconv} also follow from the argument similar to the ones given in \cite[Appendix C.7]{khalil2002nonlinear}.

\end{document}